\documentstyle[psfig,mnras_cite]{mn}
\def\lsim{~\rlap{$<$}{\lower 1.0ex\hbox{$\sim$}}}
\def\gsim{ \lower .75ex \hbox{$\sim$} \llap{\raise .27ex \hbox{$>$}} }
\def\lsim{ \lower .75ex \hbox{$\sim$} \llap{\raise .27ex \hbox{$<$}} }
\def\simprop{ \lower .75ex \hbox{$\sim$} \llap{\raise .27ex \hbox{$\propto$}} }
\title{A Comparison of Semi-Analytic and Smoothed Particle Hydrodynamics Galaxy Formation}
\author[A.~J.~Benson, F.~R.~Pearce, C.~S.~Frenk, C.~M.~Baugh and A.~Jenkins]{A.~J.~Benson$^{1,2}$, F.~R.~Pearce$^1$, C.~S.~Frenk$^1$, C.~M.~Baugh$^1$ and A.~Jenkins$^1$ \\
1. Physics Department, University of Durham, Durham DH1 3LE, England. \\
2. E-mail: A.J.Benson@dur.ac.uk \\
}

\def\FSA{FSA}
\def\SDSA{SDSA}

\begin{document}
\maketitle

\begin{abstract}
We compare the statistical properties of galaxies found in two different
models of hierarchical galaxy formation: the semi-analytic model of Cole et
al. and the smoothed particle hydrodynamics (SPH) simulations of Pearce et
al. These two techniques model gas processes very differently: by
approximate, analytic methods in the case of the semi-analytic model, and
by direct numerical integration of the equations of hydrodynamics for a set
of discrete particles in the case of SPH. Using a `stripped-down' version
of the semi-analytic model which mimics the resolution of the SPH
simulations and excludes physical processes not included in them, we find
that the two models produce an ensemble of galaxies with remarkably similar
properties, although there are some differences in the gas cooling rates
and in the number of galaxies that populate halos of different mass. The
full semi-analytic model, which has effectively no resolution limit and
includes a treatment of star formation and supernovae feedback, produces
somewhat different (but readily understandable) results. Our comparison
demonstrates that, on the whole, SPH simulations and semi-analytic models
give similar results for the thermodynamic evolution of cooling gas in
cosmological volumes. Agreement is particularly good for the present-day
global fractions of hot gas, cold dense (i.e. galactic) gas and uncollapsed
gas, for which the SPH and stripped-down semi-analytic calculations differ
by at most 25\%. In the most massive halos, the stripped-down semi-analytic
model predicts, on the whole, up to 50\% less gas in galaxies than is seen
in the SPH simulations. The two techniques apportion this cold gas somewhat
differently amongst galaxies in a given halo. This difference can be
tracked down to the greater cooling rate in massive halos in the SPH
simulation compared to the semi-analytic model. The galaxy correlation
functions in the two models agree to better than about 50\% over most pair
separations and evolve with redshift in very similar ways. Our comparison
demonstrates that these different techniques for modelling galaxy formation
produce results that are broadly consistent with each other and highlights
areas where further study is required.
\end{abstract}

\begin{keywords}
methods: numerical - galaxies: formation: kinematics and dynamics -
cosmology: theory - hydrodynamical simulation
\end{keywords}

\section{Introduction}

The properties of galaxies in the Universe are determined by the
behaviour of both the dark matter and the baryonic material from which
they are made. The dynamics of the dark matter, which are determined
by gravity alone, are now reasonably well understood. N-body
simulations (e.g. \pcite{defw}) provide an accurate description of the
evolution of structure into the highly non-linear regime where dark
matter halos form (see, for example
\pcite{arj98,col98}). Analytically, the Press-Schechter theory
\cite{PS74} predicts to within $\sim 50\%$ the distribution of halo
masses found in N-body simulations for a specified cosmology, whilst
theoretically motivated fitting functions do even better
\cite{st,smt,arj00}. Extensions of this theory \cite{BKCE,rgb91,lc93}
predict, with reasonable accuracy, the hierarchical build-up of halos
through the mergers of smaller progenitors (see, for example,
\pcite{lc94,rsetal98}). The behaviour of the baryonic matter, on the
other hand, is less well understood. The dynamics of the gas are not
determined by gravity alone but also by hydrodynamical forces and
radiative processes. Since gas must cool into dense lumps before it
can turn into stars, these processes are crucial for galaxy formation
(see, for example, \pcite{binney77,ro77,silk77,sdmw78}).

In this paper, we compare the outcome of two widely used techniques
for modelling the behaviour of gas as it forms into galaxies:
semi-analytic modelling and direct simulation using smoothed particle
hydrodynamics (SPH).  Semi-analytic models applied to cold dark matter
(CDM) cosmologies
\cite{cole91,wf91,laceysilk91,kwg93,coleetal94,rs98,coleetal99} have
met with considerable success in explaining many of the observed
properties of the galaxy population, such as the luminosity function,
the distributions of colour and morphological type, the counts as a
function of magnitude and redshift
(e.g. \pcite{wf91,lacey93,kgw94,coleetal94,gk95,b96dummy,gkdummy,kcdummy}),
the properties of Lyman-break galaxies \cite{baugh98a,governato98},
and the clustering of galaxies
\cite{kns,ketaldummy,diaferio99,baugh99,ajbdummy}.

The SPH technique \cite{L77,GM77} has been used by many authors to
model galaxy formation
(e.g. \pcite{katz91,katz92,nw93,evrard94,SMdummy,katz96,fews,weil98,navarro99,pearce99}).
These simulations have been successful in producing objects with
approximately the mass of galaxies and, in cosmological simulations,
with approximately the right abundance. However, to date no simulation
has been able to produce a realistic, rapidly rotating spiral galaxy
starting from cosmological initial conditions
(e.g. \pcite{navarro99}).

Both techniques require a number of simplifying assumptions in order
to model the evolution of cooling gas. For example, semi-analytic
models assume that dark matter halos and their associated gas
component are spherically symmetric, and that gas is efficiently
shock-heated when halos collapse. SPH, on the other hand, assumes that
gas is well represented by a set of discrete particles. The two
methods have different strengths and limitations. Semi-analytic
modelling can follow a large dynamic range of scales and is
sufficiently flexible that the effects of varying assumptions and
parameter values can be readily explored. SPH, on the other hand, does
not impose any restrictions on geometry and solves directly the
approximate evolution equations for gravitationally coupled dark
matter and dissipative gas. Limited resolution, however, restricts the
accessible dynamic range and the expense of large simulations makes it
impractical to carry out extensive parameter space explorations. In
both approaches, a phenomenological model for star formation and
feedback must be coupled to the evolution of dark matter and gas in
order to calculate observable properties of galaxies. Generally, such
models are more easily implemented in semi-analytic models than in SPH
simulations in which the behaviour of the phenomenological model
itself often depends on resolution.

The main aim of this paper is to determine the extent to which the two
techniques of semi-analytic modelling and SPH simulation produce
consistent results for the evolution of cooling gas in the cosmological
setting relevant to galaxy formation. There are several ways in which such
a comparison might be carried out. In this paper, we adopt the statistical
approach of comparing the properties of {\it populations} rather than of
individual objects. This comparison is motivated by two
considerations. Firstly, we wish to understand how the different
approximations inherent in the two techniques translate into differences in
the {\it average} properties of the two models. Secondly, both techniques
have been used (and continue to be used) to study statistical properties
such as galaxy luminosity functions and spatial correlation functions. It
is clearly important to test how reproducible these bulk properties are
using these rather different modelling techniques. A complementary approach,
which we defer to a later paper, is to compare the properties of individual
objects modelled using the two techniques. A secondary aim of our paper is
to assess how the neglect of sub-resolution processes in SPH simulations,
i.e. star formation and feedback, can affect the properties of objects
above the resolution limit.

When differences between the results of the two modelling techniques do
arise, it is difficult to know which of the two, if either, is giving the
``correct" answer. In some cases, however, we can explore the reasons for a
disagreement by altering parameters in the semi-analytic model which
describe a specific physical process, for example, the galaxy merger
timescale. In this way, we can identify specific areas of disagreement in
which further theoretical work is required.

This paper is laid out as follows. In
\S\ref{sec:moddets}, we briefly describe the SPH and
semi-analytic models, and discuss in greater detail their specific
implementation in this work. In \S\ref{sec:compare}, we compare several
properties of the galaxies calculated using the two techniques, and uncover
the reasons for some of the differences by varying key parameters in the
semi-analytic models. Finally, in \S\ref{sec:discuss}, we present our
conclusions.

\section{The SPH Simulations and Semi-analytic Models}
\label{sec:moddets}

We now present the models of galaxy formation employed in this paper. Since
the SPH and semi-analytic techniques are described in detail elsewhere, we
give only a brief overview here, referring the reader to the appropriate
references for further details where applicable.

\subsection{SPH Simulations}

\subsubsection{Techniques}

The SPH technique is a Lagrangian method in which the gaseous
component of the universe is described by a set of tracer gas elements
represented by particles within the simulation volume. Estimates of
local gas properties (and their spatial derivatives) for each particle
are derived by smoothing over the properties of the $N_{\rm SPH}$
nearest neighbour particles (see \scite{Mon92} for a review). For
simulations of galaxy formation, the gas must also be able to cool
radiatively.

The simulation volume is initially populated with dark matter and gas
particles, with a spatial distribution derived from a cosmological power
spectrum. The equations of gravity and hydrodynamics are then solved
over a succession of small timesteps in order to propagate the particle
distribution forwards in time until the present day.

\subsubsection{Simulation Specifics}
\label{sec:simspecs}

Simulations were carried out for two of the cold dark matter
cosmological models studied by \scite{arj98}: $\Lambda$CDM (mean
mass density parameter, $\Omega _0 = 0.3$; cosmological constant, in
units of $3H_0^2/{\rm c}^2$, $\Lambda _0= 0.7$; Hubble constant, in
units of 100 km s$^{-1}$ Mpc$^{-1}$, $h = 0.7$; and rms linear
fluctuation amplitude in $8h^{-1}$Mpc spheres, $\sigma_8=0.9$), and
SCDM ($\Omega _0 = 1.0$, $\Lambda _0 = 0$, $h = 0.5$, $\sigma _8 =
0.6$). The baryon fraction in each cosmology was set, from
nucleosynthesis constraints, to be $\Omega _{\rm b} h^2 = 0.015$
\cite{copi95}. The simulations, which were carried out using the
parallel AP3M-SPH code of \scite{fp97}, have $128^3$ particles of
each species in boxes of side 70$h^{-1}$ Mpc and 50$h^{-1}$ Mpc in the
$\Lambda$CDM and SCDM models respectively. The gas mass per particle
is therefore $1.4 \times 10^9$ and $1.0 \times 10^9h^{-1}{\rm
M}_\odot$ in the $\Lambda$CDM and SCDM cosmologies respectively. Since
we adopt $N_{\rm SPH}=32$, the smallest resolved objects have a gas
mass of $4.5 \times10^{10}$ and $3.2\times10^{10}h^{-1}{\rm M}_\odot$
in the two cosmologies. An unevolving gas metallicity of 0.3 times the
Solar value was assumed, taking into account that gas cooling in
objects above the resolution threshold will already have been
processed by several previous levels of the merger hierarchy. The SPH
simulations use a cooling function which is a series of power-law fits
to the results of \scite{raysmith}. If instead the tabulated
cooling function of \scite{suthdop} is used (this being the
function used in the semi-analytic models) then around 10\% more cold
gas results by $z=0$ \cite{kay99}.

We employed a comoving $\beta$-spline gravitational softening
equivalent to a Plummer softening length of $35h^{-1}$kpc for $z>2.5$
in $\Lambda$CDM and $25h^{-1}$kpc for $z>1.5$ in SCDM. At lower
redshifts, the softening remained fixed at $10h^{-1}$kpc in physical
coordinates, and the minimum SPH spatial resolution was also set to
match this value. Approximately 10,000 timesteps were used in each
simulation to evolve from $z=50$ to $z=0$. With our chosen parameters,
the simulations are able to follow the cooling of gas into galactic
dark matter halos. The resulting ``galaxies'' typically have 50-1000
particles.

Dark matter halos were identified at $z=0$, 0.1, 0.2, 0.3, 0.5, 1,
2 and 3. Simulation outputs at higher redshifts were available,
but so few galaxies have formed beyond $z=3$ in these small volumes
that no statistically meaningful comparison could be made at higher
redshifts. Halos were found using the friends-of-friends (FOF) group
finding algorithm \cite{defw} with a linking length of $b=0.2$ times
the mean interparticle separation in SCDM. In $\Lambda$CDM, we assume
halos form with the overdensity predicted by the model of
\scite{eke96} and therefore use a linking length which grows from
$b=0.164$ at $z=0.0$ to $b=0.200$ at $z=3$. Consistent definitions
of halo overdensities are adopted in the semi-analytic models
considered.

``Galaxies'' were also found using this algorithm, but with a much
smaller linking length of $b=0.0168$. Galaxies are made of gas
particles that have cooled below 12,000~K and represent overdensities
of $\gsim 100,000$. A complete description of the simulations and the
properties of the simulated galaxies is given by
\scite{pearce2000}.

\subsubsection{Assumptions and Limitations}
\label{sec:SPHass}

The key assumption of the SPH technique is that the evolution of gas
may be approximated by the evolution of a set of particles. Each
particle may be thought of as a packet of gas that ``carries'' with it
the thermodynamical properties of the system.

The smoothing inherent in the SPH technique introduces problems
whenever gas properties vary discontinuously (or at least on scales
much smaller than the smoothing scale). In the case of shocks, an
artificial viscosity term is used to capture the shock and prevent it
from being smoothed away by the SPH algorithm. Another example of this
kind of problem occurs in a multiphase gas (although see
\pcite{thomas00}), in which the sharp boundary between phases is
smoothed over, causing the phases to diffuse into one another. This
problem can lead to runaway cooling in the centres of dark matter
halos, as happened, for example, in one of the simulations of
\scite{fews}. The simulations of \scite{pearce99} attempt to
circumvent this problem by ignoring the contribution of cold
($T<12,000$K) particles in the computation of the densities of hot
($T>10^5$K) particles. (For a complete discussion of this
approximation see \pcite{thacker99}) An important consequence of this
approximation is that galaxies in the simulations match the shape of
the observed galaxy luminosity function at the bright end.

A further limitation arises from the fact that the sizes of the
galaxies that form in the simulations are determined primarily by the
gravitational force softening length rather than by any real physical
process. This raises the possibility of enhanced tidal disruption,
drag, and merging within dark matter halos. Whilst the softening
length is kept fixed in physical coordinates at low redshift, it is
fixed in comoving coordinates at high redshift, as described in
\S\ref{sec:simspecs}. Thus for $z>1.5$ the {\it physical} softening length
in the $\Lambda$CDM simulation is larger than in the SCDM simulation and,
as a result, non-physical effects due to softening may be expected to
be more pronounced at early times in the $\Lambda$CDM simulation.

Although a variety of prescriptions have been tried in attempts to
model supernovae feedback in SPH simulations, usually by converting
cold gas into ``star particles'' which then inject thermal and kinetic
energy into the surrounding gas (e.g. \pcite{nw93,SM95,katz96}), this
process remains poorly understood. In cosmological SPH simulations,
gas can only begin to cool efficiently in objects well above the
minimum resolved halo mass, around several times $10^{11}h^{-1}{\rm
M}_\odot$ in our case. Thus, resolution effects prevent all the gas
from cooling in small halos at high redshift, a process that, in
reality, is probably due to feedback from supernovae or other
energetic sources. The resolution of our simulation was, in fact,
chosen to ensure that the fraction of baryons that cools by the
present in the SCDM model is comparable to the observed fraction of
cold gas and stars in galaxies today. This was achieved by carrying
out several test simulations with varying resolution until the desired
cold gas fraction was obtained \cite{kay99}.

\subsection{Semi-analytic models of galaxy formation}

\subsubsection{Techniques}

\begin{figure*}
\hspace{3mm}\psfig{file=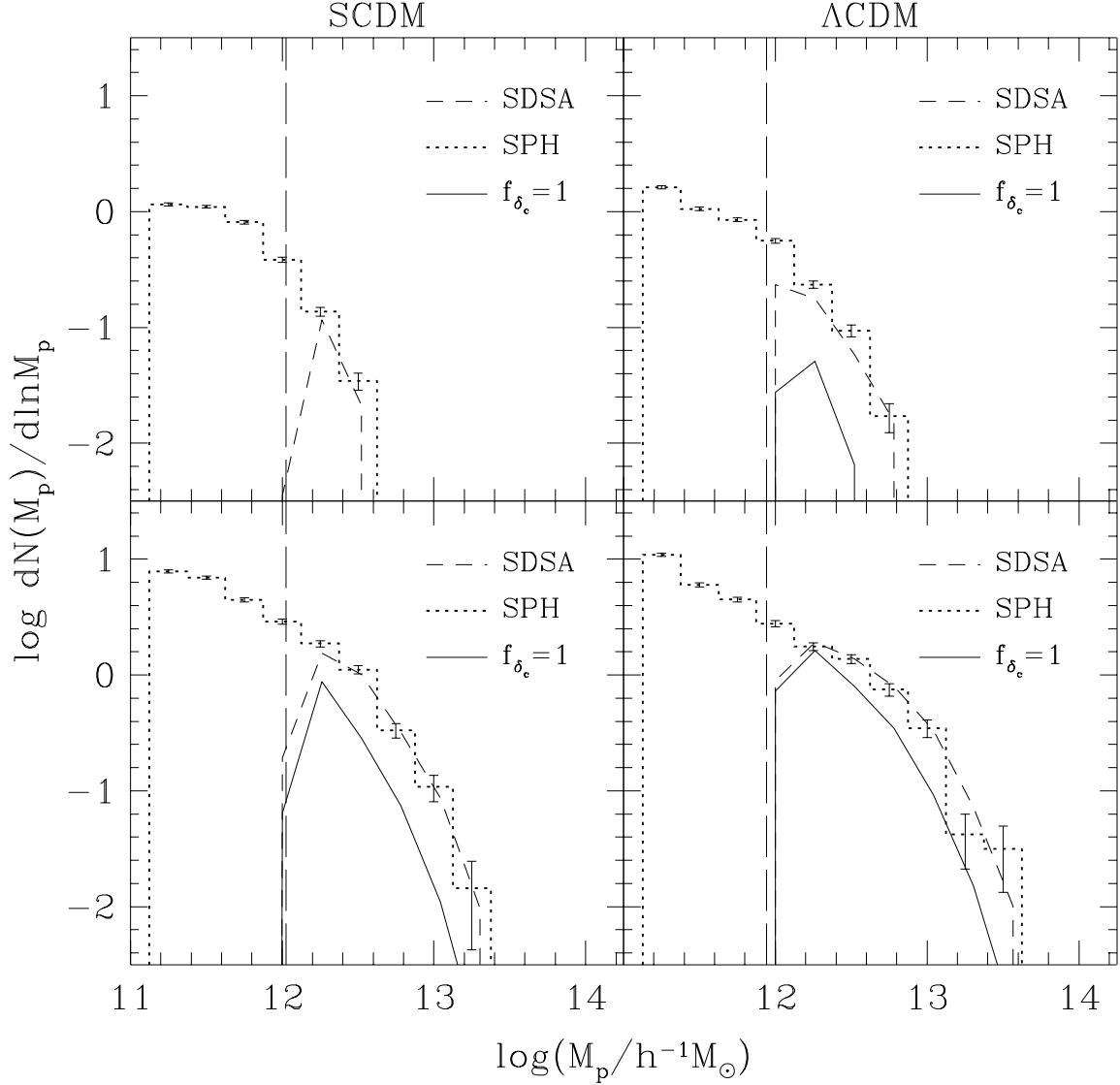,width=160mm,bbllx=0mm,bblly=85mm,bburx=190mm,bbury=270mm,clip=}
\caption{The mass function of dark matter halos at $z=2$ which are
progenitors of present-day halos of mass $10^{12}$--$10^{13}h^{-1}{\rm
M}_\odot$ (upper panels) and $10^{13}$--$10^{14}h^{-1}{\rm M}_\odot$
(lower panels) in the SCDM (left-hand panels) and $\Lambda$CDM
(right-hand panels) cosmologies. The dotted histograms are the mass
functions in the SPH simulations. The thin solid lines show the mass
functions obtained from the \SDSA\ models using the standard
definition of the extrapolated critical linear overdensity for
collapse from the spherical collapse model (i.e. with the parameter
$f_{\delta_{\rm c}}=1$), whilst the dashed lines show the \SDSA\ model
result when $f_{\delta_{\rm c}}$ is chosen to give a good match to the
progenitor mass function in the SPH simulations
(c.f. eqn.~\protect\ref{eq:fdc}). Vertical dashed lines show the
resolution limit imposed on the \SDSA\ model merger trees.}
\label{fig:DMprog}
\end{figure*}

In semi-analytic models, some of the processes involved in galaxy
formation (e.g. the growth of dark matter halos by mergers of smaller
halos) are followed using analytic solutions and Monte-Carlo
techniques. Other, more uncertain processes, such as feedback from
supernovae are modelled by means of simple, physically motivated
rules. Typically, each such rule contains one or two free parameters
which are constrained using observations of galaxies in the local
Universe (e.g. \pcite{coleetal99}).

In semi-analytic models, the dynamics of the gas are strongly coupled
to the evolution of dark matter halos and to the processes of star
formation and feedback. The starting point for our own modelling is
the set of dark matter halos at $z=0$ drawn from the Press-Schechter
mass function. A merging history for each halo is then constructed
using the extended Press-Schechter formalism. Beginning with the
earliest progenitor halo, our model assumes that gas (initially
assumed to have zero metallicity) is shock-heated to the virial
temperature of the halo, after which it begins to cool according to a
specified cooling function. Any gas that does cool forms a galaxy at
the centre of the halo within which stars begin to form at a specified
rate, producing both metals and supernovae. Supernovae reheat some of
the gas in the galaxy, ejecting it back out into the surrounding
halo. (This gas is not allowed to cool again until the halo has merged
to form part of a larger halo.) These processes continue until the
halo mass has increased by a factor of two or more, either by merging
with a larger halo, or by numerous accretions of smaller halos. Any
leftover hot gas becomes part of the new halo, and the largest galaxy
of the newly formed halo becomes the central galaxy, onto which
further gas can cool. Any other galaxies become satellites in the new
halo, and may eventually merge with the central galaxy due to energy
loss by dynamical friction. The full model includes other processes
such as stellar population synthesis and morphological evolution which
are not directly relevant to the present work.

As well as this ``full'' semi-analytic (\FSA) model, for the purposes of
this work we have constructed a ``stripped-down'' semi-analytic (\SDSA)
model which is designed to be directly comparable to the SPH simulations
(in a statistical sense). In the \SDSA\ model, we switch off star formation
and the associated supernovae feedback and chemical enrichment, since these
processes are not included in the SPH calculation. Instead, we assume a
fixed metallicity of 0.3 times the Solar value, just as in the
simulations. We also mimic the SPH resolution by truncating halo merger
trees at $N'_{\rm SPH}$ times the dark matter particle mass in the
simulation, and by switching off gas cooling when the hot gas mass is less
than $N'_{\rm SPH}$ times the gas particle mass. The parameter $N'_{\rm
SPH}$ is set to $2\times N_{\rm SPH}=64$, i.e. twice the number of
particles in the SPH smoothing kernel. This value was chosen since it
allows the \SDSA\ model to match the position of the peak in the galaxy
mass function in the SPH simulations (as will be shown in
\S\ref{sec:massfunc}, see Fig.~7). The \FSA\ model has no such
truncation of merger trees and has effectively unlimited resolution
(in practice, we resolve progenitor halos down to masses several
hundred times smaller than in the SPH and \SDSA\ models).

To further emulate conditions in the SPH simulations in the \SDSA\ model,
we replace the Press-Schechter formula for the mass function of dark matter
halos (which is commonly used in semi-analytic models, including our \FSA\
model) with the formula proposed by \scite{smt} which provides a
better match to the results of large N-body simulations \cite{arj00}.
Although this ensures that the abundance of halos at $z=0$ in the \SDSA\
model is similar to that in the SPH simulations, it does not guarantee that
the distribution of progenitor halo masses will also be the same. In fact,
as has been shown by
\scite{rsetal98}, we find that at high redshifts the merger trees in
our \SDSA\ model underpredict the abundance of progenitor halos seen in
the SPH simulations. We reconcile the progenitor distributions in the two
models by adopting a simple, empirical correction. We scale the
value of the extrapolated critical linear threshold for collapse from the
spherical collapse model, $\delta_{\rm c}$, by a redshift-independent
factor, $f_{\delta_{\rm c}}$, such that $\delta^{\rm eff}_{\rm c}
=f_{\delta_{\rm c}}\delta_{\rm c}$ . (Thus, in the standard form of the
extended Press-Schechter theory, $f_{\delta_{\rm c}} = 1$.) This approach
is similar to that proposed by \scite{tormen98}. The factor
$f_{\delta_{\rm c}}$ is allowed to be a function of the $z=0$ halo mass. We
find that the following simple form for $f_{\delta_{\rm c}}$ provides a
reasonable fit to progenitor mass functions in the SPH simulations over the
redshift range 0 to 3:
\begin{equation}
f_{\delta_{\rm c}} = \left\{ \begin{array}{ll} 1 +
0.09 \left[ \log (M_{\rm halo}/h^{-1}{\rm M}_\odot) - 16.56\right] , &
\hbox{SCDM} \\ 1 + 0.14 \left[ \log (M_{\rm halo}/h^{-1}{\rm M}_\odot) -
15.64\right] , & \Lambda\hbox{CDM.} \end{array} \right.
\label{eq:fdc}
\end{equation}
This correction to the extended Press-Schechter theory is based purely
on an empirical fit to our two simulations. It is designed to minimize
differences in the statistical properties of the dark halos in the two
models, so that we can focus on differences in their gasdynamical
properties. Given the limited statistics provided by our relatively
small simulation volumes, this correction should not be regarded as
appropriate in a general sense and may not be accurate for larger
volumes or for different power spectra to those considered here. In
Fig. \ref{fig:DMprog} we show the mass function of progenitor halos at
$z=2$ for parent halos in the mass ranges
$10^{12}$--$10^{13}h^{-1}{\rm M}_\odot$ and
$10^{13}$--$10^{14}h^{-1}{\rm M}_\odot$, defined so that $({\rm
d}N(M_{\rm P})/{\rm d}\ln M_{\rm P}){\rm d}\ln M_{\rm P}$ is the
number of progenitors in the mass range $\ln M_{\rm P} $ to $\ln
M_{\rm P} +{\rm d}\ln M_{\rm P}$ per parent halo. The dashed lines
indicate the distribution from our \SDSA\ model with the above
correction, whilst the solid lines show the model without the
correction. The correction succeeds in ensuring that the mass
functions of progenitor halos are statistically similar in the \SDSA\
and SPH models.

\subsubsection{Specifics}
\label{sec:params}

The semi-analytic model of \scite{coleetal99} was used to simulate
galaxy formation in a large sample of dark matter halos spanning a wide
range in mass. Specifically, for the \SDSA\ model we simulated halos in 28
mass bins spaced uniformly in the logarithm of halo mass between $10^{11}$
and $10^{15}h^{-1}{\rm M}_\odot$. For each mass bin, 100 halos were simulated. In
the \FSA\ model the mass range was extended down to $10^{10}h^{-1}{\rm M}_\odot$,
and between 4 and 20 halos were simulated in each mass bin (fewer halos
were simulated for the most massive bins as these are computationally more
expensive). To study the clustering of galaxies in the semi-analytic
models, we used the techniques of \scite{ajbdummy} and populated halos
in the SPH simulations with ``semi-analytic'' galaxies.

The cosmological parameters required as input into the semi-analytic
model ($\Omega _0$, $\Lambda_0$, $h$, $\sigma _8$, $\Omega _{\rm b}$)
were set to the same values used in the SPH simulations. The remaining
parameters of the \FSA\ model were set equal to the values chosen by
\scite{coleetal99} (for $\Lambda$CDM) and by \scite{ajbclust} (for
SCDM, for which we used the parameters of their $\tau$CDM
model). Parameters were chosen so as to obtain a model which produces
a reasonable match to the local B and K-band luminosity functions and
other local data (as described by \pcite{coleetal99}). These
parameters can be split into two classes: those that affect the
results of the \SDSA\ model and those that do not. The latter,
however, are still important for specifying the behaviour of the \FSA\
model. Parameters that do not affect the \SDSA\ model are those which
govern star formation, feedback from supernovae and the production of
metals by stars. We do not discuss them in any detail here but refer
the reader to \scite{coleetal99} for a full description.

The parameters which do affect the results of the \SDSA\ model (and whose
specific values tend to be inspired by the results of simulations) are the
following: (i) $f_{\mathrm df}$, the dynamical friction coefficient that
determines the merger timescale for galaxies orbiting in halos (see
eqn. \ref{eq:merge}); (ii) the dark matter density profile; (iii) the gas
density profile; (iv) the progenitor halo mass resolution; and (v) the
critical mass for cooling ((iv) and (v) are both specified by $N'_{\mathrm
SPH}$). Unless otherwise stated, we set $f_{\rm df}=1$, $N'_{\rm SPH}=64$,
and assume (a) that the dark matter density profile has the form proposed
by \scite{nfwdummy}, namely,
\begin{equation} \rho
(r) \propto {1 \over r/r_{\mathrm s} (1+[r/r_{\mathrm s}]^2)},
\end{equation}
where $r_{\mathrm s}$ is a scale-length, and (b) that the gas density
has an isothermal profile at large radii, and a constant density core
of size $r_{\rm c}$, i.e.
\begin{equation} \rho (r) \propto {1 \over r^2 + r_{\mathrm c}^2}.
\end{equation}
The core radius is initially set to some fraction of the NFW
scale-length, $r_{\rm s}$, of the dark matter halo. Our standard
choice for this fraction is 0.33, motivated by the results of
hydrodynamical simulations of cluster formation
\cite{jfn95,eke98}. The gas that is able to cool in a halo is the
densest gas, which has the lowest entropy. When halos merge to form a
new halo this low entropy gas, which would normally settle into the
inner parts of the halo, is missing. We take this into account by
increasing the core radius of later generations of halos so that the
gas density at the virial radius is the same as it would have been if
no gas had cooled in progenitors (for a full description see
\pcite{coleetal99}). The model also allows us the option of keeping
the core radius fixed, which we explore below.  The inclusion of a
core in the hot gas profile prevents the formation of extremely bright
galaxies in the centres of groups and clusters, which would otherwise
lead to a disagreement with the shape of the bright end of the
observed galaxy luminosity function. The reasons for choosing this
particular profile are therefore identical in spirit to those for
preventing runaway cooling in the SPH simulations (see
\S\ref{sec:SPHass}).

\subsubsection{Assumptions and Limitations}

Semi-analytic models make several assumptions in the treatment of gas
in order to obtain simple, analytic solutions to complex
hydrodynamical processes. We have already mentioned the important
assumptions of spherical symmetry and of the shock-heating of the gas to the
virial temperature of its associated halo. The hot gas is then further
assumed to settle into a distribution with a universal form. Finally,
the amount of gas that is able to cool by time $t$ after the formation
of the halo is identified with the gas contained within the radius at
which the cooling time equals $t$. Once it has cooled, this gas is
assumed to flow to the centre of the halo, where it is available for
star formation, provided that the free-fall time for the gas is also
less than $t$. We shall refer to this as the ``cooling radius''
prescription.

\section{Comparison of the two models}
\label{sec:compare}

In this section we compare several properties of the galaxy
populations that form in our models and consider how this comparison
is affected by varying certain assumptions and parameter values.

\subsection{Properties of halo gas}

We begin by comparing the most basic quantities calculated by each
technique, namely the fraction of gas in the hot and cold phases, both
globally and as a function of dark matter halo mass. For these
purposes, we define a `hot halo gas phase' as gas hotter than $10^5$K;
a `galaxy phase' represented by cool, dense gas in the SPH simulation
and SDSA, and also including stars in disks and spheroids in the
FSA; and an `uncollapsed gas phase' consisting of everything else ---
i.e. gas outside virialised halos. Note that for the galaxy phase, we
consider only galaxies with a mass greater than $N'_{\rm SPH}$ gas
particles in the SPH and \SDSA\ models, but include galaxies of all
masses in the \FSA\ model.

\begin{figure*}
\hspace{2.5mm}\psfig{file=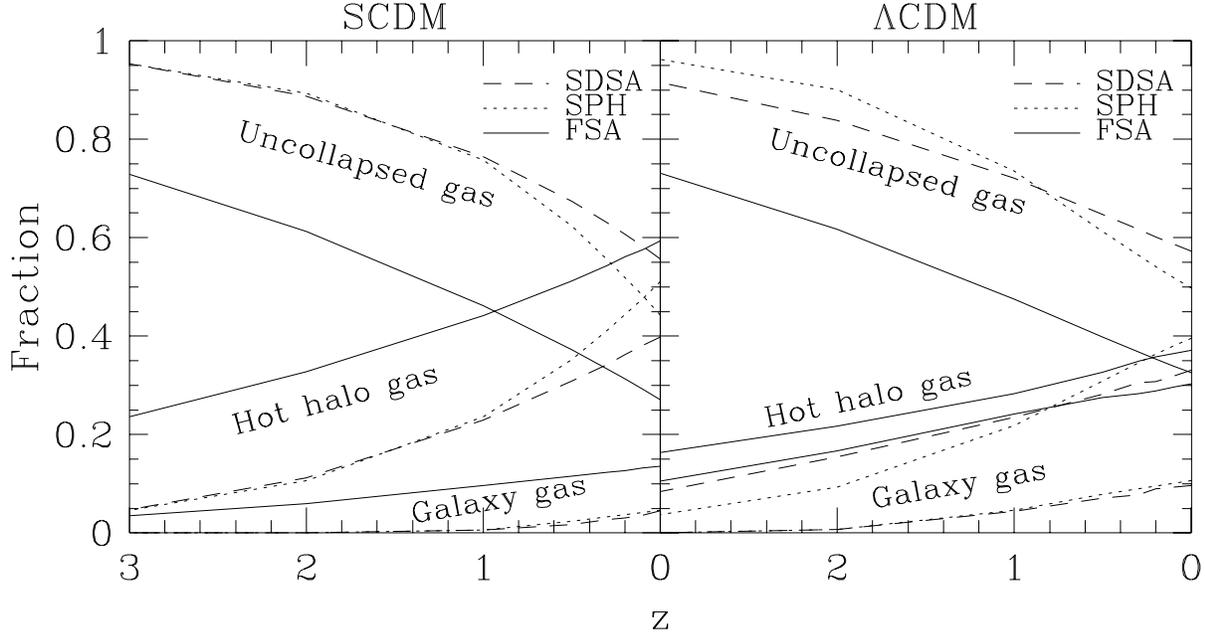,width=160mm,bbllx=5mm,bblly=170mm,bburx=187mm,bbury=270mm,clip=}
\caption{The global fraction of gas in each of three phases: hot
halo gas, uncollapsed gas and galaxy gas. Dotted lines correspond
to the SPH simulation,dashed lines to the \SDSA\ model, and solid
lines to the \FSA\ model.} \label{fig:gastemp}
\end{figure*}

In the Press-Schechter (or Sheth-Mo-Tormen) theory, all the matter in the
universe is deemed to be in halos of some mass, and semi-analytic models
assume that gas in halos is shock heated to the halo virial temperature. We
can therefore determine the fraction of gas in the uncollapsed gas phase in
the \FSA\ model simply by integrating over the analytical mass function
\cite{smt} from zero mass to the mass corresponding to a virial
temperature of $10^5$K. According to the spherical top-hat model of halo
formation, the mass corresponding to $10^5$K is:
\begin{equation}
M_{\rm 10^5{\rm K}} = 3.5\times 10^{10} (1+z)^{-3/2} \Omega_0^{-1/2} \left( {200 \over \Delta
_{\rm c}(z)} \right)^{1/2}
h^{-1} M_{\odot},
\end{equation}
where $\Delta _{\rm c}(z)$ is the overdensity of a newly formed,
virialised dark matter halo at redshift $z$
(e.g. \pcite{eke96}). Since some halos hotter than $10^5{\rm K}$ are
not resolved in the \SDSA\ model, the integration in this case is
carried out from zero mass to $M_{10^5{\rm K}}$ or to $N'_{\rm SPH}$
dark matter particle masses, whichever is largest. This estimate does
not correspond exactly to the situation in the SPH simulation in which
the largest halos are surrounded by gas at temperatures above $10^5$K
which extends beyond the virial radius. Because of this, the \SDSA\
model calculation will effectively overestimate the amount of
uncollapsed gas relative to the SPH simulation. On the other hand, gas
in the SPH simulation tends to be slightly more extended than assumed
in the semi-analytic model (i.e. the simulated clusters tend to have a
baryonic content slightly smaller than the universal baryon fraction
within a radius enclosing an overdensity of 200 -- see
e.g. \pcite{frenk99}). These two effects counteract each other to some
degree.

The amount of gas in the hot and galaxy phases depends upon the rate at
which gas cools. Therefore this comparison tests model assumptions relating
to the process of gas cooling, such as spherical symmetry and the cooling
radius prescription in the semi-analytic models or the effects of smoothing
in SPH. This test will therefore be sensitive to the choice of gas density
profile in the semi-analytic models and to $N_{\rm SPH}$ in the
simulations. Since this comparison is concerned only with the total amount
of gas in different phases, it is insensitive to the way in which the gas
is apportioned into galaxies within a single halo, at least in the SPH and
SDSA models. In the \FSA\ model, some dependence on galaxy merger rates may
exist, since merging can affect the star formation rate in a galaxy and
thus alter the amount of gas reheated by feedback, as well as the rate of
chemical evolution, which in turn alters the cooling rates in subsequent
generations of halos.

\subsubsection{Global gas fractions}
\label{sec:global}

Figure \ref{fig:gastemp} shows the fraction of gas in each of the
three phases: hot halo, galaxy and uncollapsed, as a function of
redshift. In both cosmologies, the uncollapsed gas fraction in the
\SDSA\ model is quite close to, although somewhat larger (by $\lsim
0.1$), than in the SPH simulation at low redshifts. At $z=0$, the
fractional difference is $\lsim 30\%$. Given the caveats mentioned above,
this level of agreement is pleasing. In the \FSA\ model, the fraction of
gas in the uncollapsed phase is significantly lower than in the SPH
simulation and the \SDSA\ models. The only differences between the \FSA\
and \SDSA\ models in this calculation is the numerical resolution and the
use of the Press-Schecther mass function in the \FSA\ model and the
Sheth-Mo-Tormen in the \SDSA\ model. It turns out that these differences
contribute about equally to the discrepancy in the fraction of uncollapsed
gas in the two cases at $z=0$. (The Press-Schechter mass function contains
more low temperature halos than the Sheth-Mo-Tormen mass function.) At
higher redshift, resolution effects are the dominant factor. The gas
belonging to sub-resolution halos is classed as uncollapsed gas in the
\SDSA\ and SPH cases, but is accounted for as hot halo or galaxy gas in the
\FSA\ case.

\begin{figure}
\psfig{file=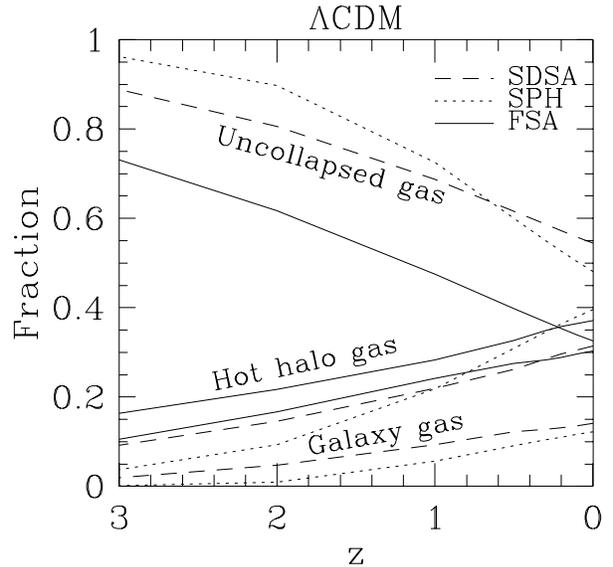,width=80mm,bbllx=85mm,bblly=170mm,bburx=187mm,bbury=270mm,clip=}
\caption{The global fraction of gas in each of three phases: hot
halo gas, uncollapsed gas and galaxy gas. Results are shown for
the $\Lambda$CDM cosmology with $N'_{\rm SPH}=32$. Dotted lines
correspond to the SPH simulation, dashed lines to the \SDSA\
model, and solid lines to the \FSA\ model.}
\label{fig:gastemp_NSPH32}
\end{figure}

\begin{figure*}
\hspace{3mm}\psfig{file=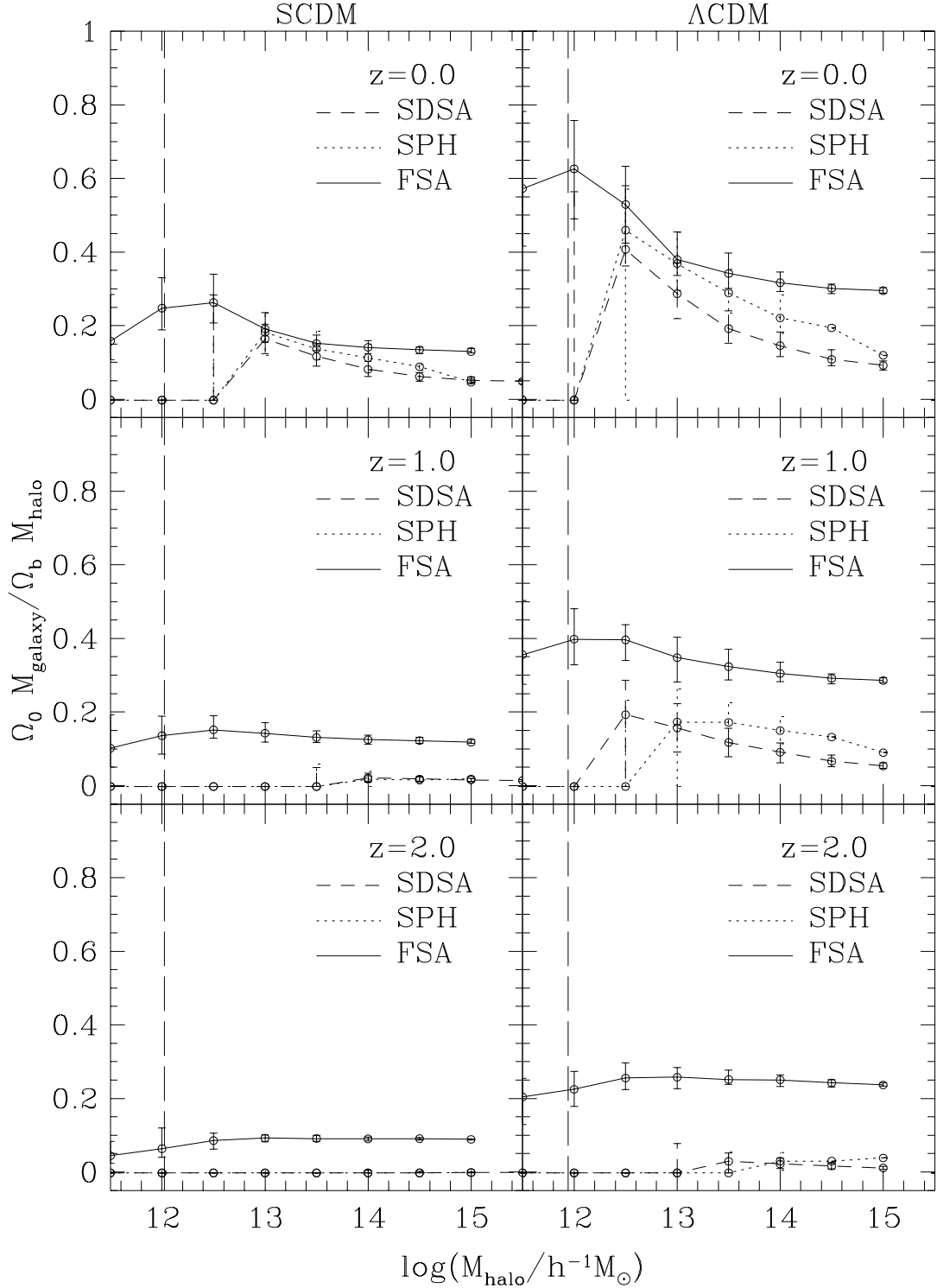,width=160mm} \caption{The mass of
gas relative to the total halo mass (i.e. dark matter plus gas) at
$z=0$ in progenitor galaxies, as a function of present-day halo
mass. (At $z=0$ we include all galaxies in the halo, whilst at higher
redshifts we include all galaxies in progenitors of this halo.) The
ordinate shows this fraction divided by the universal baryon
fraction. Only progenitor galaxies more massive than $N'_{\rm SPH}$
gas particles are considered. Results are shown at $z= 0$, 1 and 2 for
SCDM (left-hand panels) and $\Lambda$CDM (right-hand panels). The
solid lines show the median of the distribution in the \FSA\ model,
with errorbars indicating the 10 and 90 percentiles. The dotted lines
show the corresponding fraction for galaxies in the SPH simulation,
and the dashed lines for galaxies in the \SDSA\ model. The
long-dashed, vertical lines indicate the mass of those halos which, on
average, contain a total gas mass (including the hot and cold
components) equal to 64 times the SPH gas particle mass (assuming that
the gas mass is $\Omega_{\rm b}/\Omega_0$ times the total halo mass).}
\label{fig:coldgas}
\end{figure*}

\begin{figure}
\psfig{file=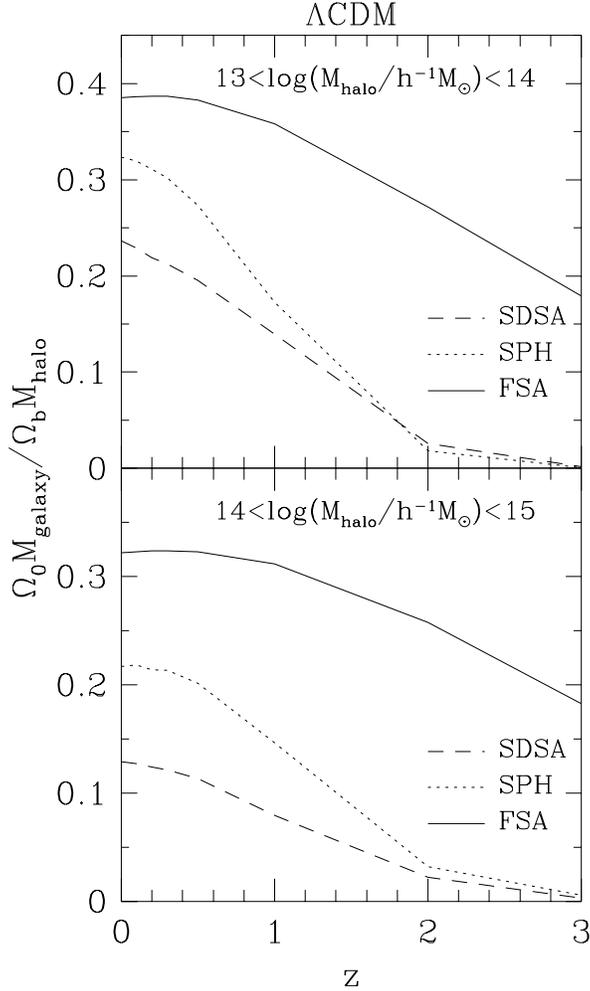,width=80mm,bbllx=5mm,bblly=90mm,bburx=110mm,bbury=270mm,clip=}
\caption{The mean mass of gas in the galaxy phase relative to the
total mass (i.e. dark matter plus gas) of the $z=0$ halo, scaled by
the universal baryon fraction, as a function of redshift in
progenitors of present-day halos of mass $10^{13}$--$10^{14}h^{-1}{\rm
M}_\odot$ (upper panel) and $10^{14}$--$10^{15}h^{-1}{\rm M}_\odot$
(lower panel). Results are shown for the $\Lambda$CDM
cosmology. Dashed lines show the \SDSA\ models, dotted lines the SPH
models and solid lines the \FSA\ models. } \label{fig:halogasevolve}
\end{figure}

\begin{figure}
\psfig{file=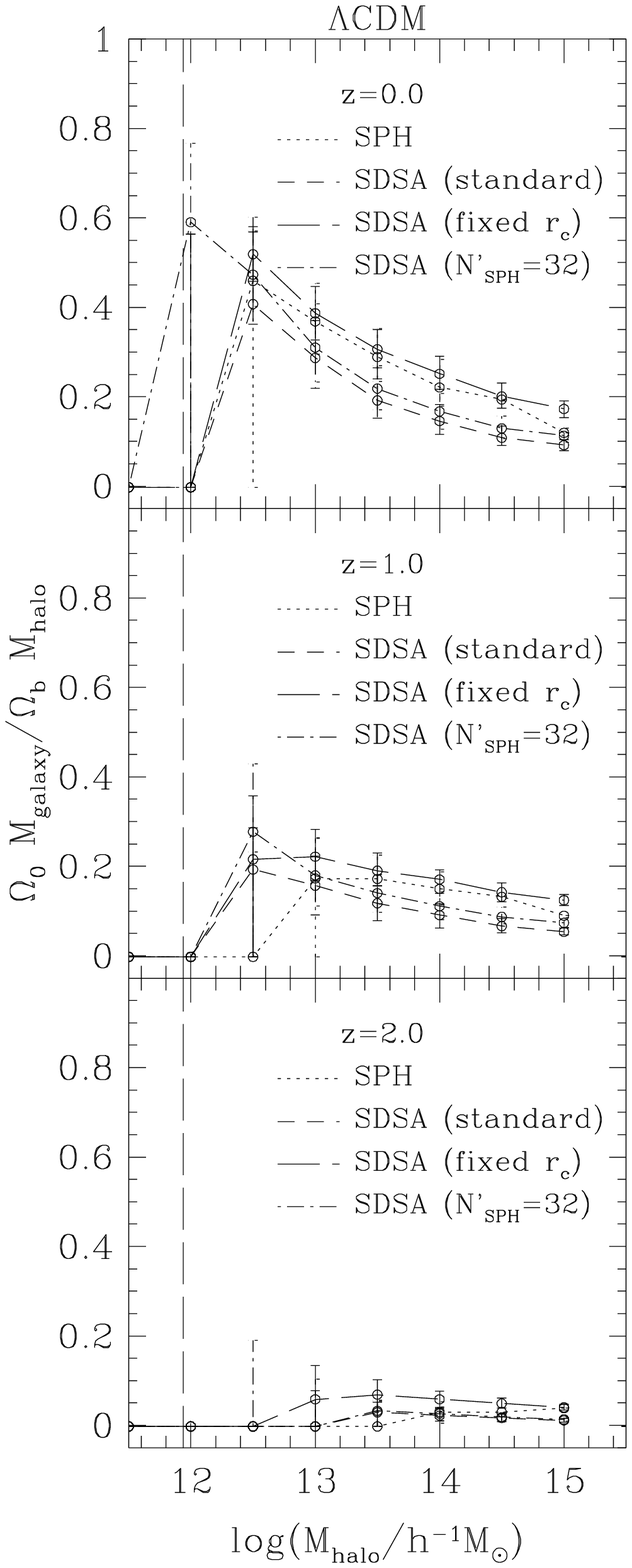,width=80mm,bbllx=0mm,bblly=10mm,bburx=110mm,bbury=270mm,clip=}
\caption{The mass of galaxy gas relative to the total halo mass
(i.e. dark matter plus gas) at $z=0$ in progenitor galaxies, as a
function of present-day halo mass in the $\Lambda$CDM cosmology. (At
$z=0$ we include all galaxies in the halo, whilst at higher redshifts
we include all galaxies in progenitors of this halo.) The ordinate
shows this fraction divided by the universal baryon fraction.  The
dotted line shows the median, and the errorbars the 10 and 90
percentiles of the distribution in the SPH simulation. The remaining
lines show predictions calculated from the \SDSA\ model for: standard
parameters (dashed line); a hot gas core radius which is fixed at 0.33
times the NFW scale length (long-dashed line); and a critical mass for
cooling equal to 32 times the SPH gas particle mass (dot-dashed
line). The long dashed, vertical line indicates the mass at which
halos contain a total mass in gas (i.e. hot halo and galaxy gas) equal
to 64 times the SPH gas particle mass (assuming that the gas mass is
$\Omega_{\rm b}/\Omega_0$ times the total halo mass). Results are
shown at $z=0$, 1 and 2 as indicated in the panels.}
\label{fig:coldgas_alt}
\end{figure}

In the galaxy phase, which is the most interesting from the perspective of
galaxy formation, the \SDSA\ and SPH models again agree extremely well at
all redshifts in both cosmologies. The fractional difference, defined as
$\left| M_{\rm SDSA}-M_{\rm SPH}\right| / M_{\rm SPH}$ where $M_{\rm SDSA}$
and $M_{\rm SPH}$ are the masses of galaxy gas in the \SDSA\ and SPH models
respectively, is close to 50\% at $z=2$ in the SCDM cosmology, but is much
smaller in $\Lambda$CDM. Half of the $z=0$ mass of galaxy gas has cooled by
$z\approx 0.5$ and $z\approx 1.0$ in SCDM and $\Lambda$CDM
respectively. Below these redshifts, the fractional difference between the
\SDSA\ and SPH models is everywhere less than 25\%. The agreement between
the hot halo gas in the \SDSA\ and SPH models is also very good. For
$\Lambda$CDM, at $z=0$, the two models differ by only 20\%. This is a
non-trivial result. One of the most uncertain assumptions of the
semi-analytic technique is that gas is shock-heated to the virial
temperature of the parent halo. It is therefore reassuring that the amount
of hot halo gas turns out to be similar to that seen in the SPH
simulations.

The \FSA\ model predicts significantly more galaxy gas than the SPH
and \SDSA\ models. Since the merger trees in this model have
effectively no mass resolution limit, gas cools very efficiently at
high redshifts in small halos. However, feedback reheats a significant
fraction of this gas, reducing the galaxy phase fraction. In SCDM,
stronger feedback is required to match the observed galaxy luminosity
function than in $\Lambda$CDM, and this is reflected in a smaller
galaxy gas fraction.

Since it is $N'_{\rm SPH}$ that determines which halos are resolved in the
\SDSA\ model, this parameter affects the gas fractions in all three
phases. For example, reducing $N'_{\rm SPH}$ from 64 to 32 worsens the
agreement between the galaxy phase fractions in the \SDSA\ and SPH models
in the $\Lambda$CDM cosmology as shown in Fig. \ref{fig:gastemp_NSPH32},
although the two models are still reasonably close, differing only by
approximately $15\%$ at $z=0$. The differences produced by this entirely
plausible change in $N'_{\rm SPH}$ are indicative of the degree of
uncertainty inherent in these comparisons.

\subsubsection{Galaxy gas fractions in halos}
\label{sec:coldgasfrac}

Figure \ref{fig:coldgas} shows the mass of ``galaxy-phase'' gas,
expressed as a fraction of the total (i.e. dark matter plus gas) halo
mass and scaled by the universal baryon fraction. At $z>0$ the
quantity shown is the fraction of gas which exists in progenitors of
the $z=0$ halos in each mass bin. This is a bulk quantity for each
halo and is independent of the way in which gas is divided among the
galaxies that reside in each halo. The \SDSA\ model predicts somewhat
less galaxy gas than the SPH simulation, particularly in the
$\Lambda$CDM cosmology. The difference is typically $\sim 50\%$,
except at low masses where the agreement is better. (We do not plot
error bars for the largest mass bins in the SPH simulations because
they contain only one or two halos.) We note that this is
significantly more than the 10\% change in cooled gas mass found by
\scite{kay99} when the \scite{suthdop} cooling function was
used in SPH simulations (which, in any case, increased the amount of
gas that cooled). At higher redshifts, the two model curves become
closer, indicating different mean rates of gas cooling. At the
smallest masses plotted, the SPH and \SDSA\ curves turn over near the
critical cooling mass (marked on the plot by the vertical long dashed
line) because of resolution effects, while the \FSA\ model turns over
because of the effects of feedback which begin to become efficient at
reheating cold galaxy gas in halos below $\sim 10^{12} h^{-1}
M_{\odot}$.

As noted in \S\ref{sec:global}, the \FSA\ model cools much more gas
into the galaxy phase than the SPH and \SDSA\ models. Much of this gas
cools at high redshift in low mass progenitor halos (which are
unresolved in the \SDSA\ and SPH models) as may be seen from the lower
panels of Fig. \ref{fig:coldgas}.

Differences in the cooling rates in the $\Lambda$CDM cosmology can be
seen more clearly in Fig. \ref{fig:halogasevolve}, where we plot the
mean fraction of cold gas in progenitor galaxies for two bins of
present-day halo mass: $10^{13}$--$10^{14}h^{-1}{\rm M}_\odot$ (upper
panel) and $10^{14}$--$10^{15}h^{-1}{\rm M}_\odot$ (lower panel). At
high redshifts, the SPH cooling rate is very similar to the cooling
rate in the \SDSA\ model, but it becomes faster at lower
redshifts. Similar trends are seen in the SCDM cosmology, although the
differences are smaller, leading to very similar present-day gas
fractions in the two models. In $\Lambda$CDM, the SPH model ends up
with 30-70\% more cool gas. In both cosmologies, the net cooling rate
in the \FSA\ model is slower than in the SPH model (particularly in the
lower range of halo mass), but comparable to that in the \SDSA\
model. Whilst the \FSA\ model resolves more progenitors, which speeds
cooling, feedback reduces the effective cooling rate. However, since
the \FSA\ model contains progenitor halos at much higher redshifts
than the other models (due to its greater resolution), more gas has
already cooled in the \FSA\ model by $z=2$ than in either the SPH or
\SDSA\ models.

Since the amount of gas that can cool depends upon the density profile
assumed for the hot halo and on the critical cooling mass, it is
interesting to see how sensitive the agreement between the \SDSA\
model and the SPH simulation is to variations in these parameters. We
show the result of this test in Fig.~\ref{fig:coldgas_alt} for the
$\Lambda$CDM cosmology. (The trends are similar for SCDM.) Halving the
critical mass required for gas cooling (i.e. reducing $N'_{\rm SPH}$
from 64 to 32) in the \SDSA\ model has a small effect, marginally
improving the agreement with the SPH simulation. (Note that for
$N'_{\rm SPH}=32$, we plot results only for objects more massive than
64 gas particles in order to compare directly to the other curves.)
The critical cooling mass is not, of course, a precise number, and
this comparison suggests that we may have been too conservative in
setting it equal to $2\times N_{\rm SPH}$. In any case, it appears
that the agreement between the galaxy gas mass as a function of halo
mass in the \SDSA\ and SPH models is better than one might have
expected.

Keeping the gas core radius fixed at $0.33r_{\rm s}$ in all halos,
rather than letting it grow as in our standard model, allows
significantly more gas to cool in the \SDSA\ model. This brings this
model into excellent agreement with the SPH model at $z=0$, as may be
seen in Fig.~\ref{fig:coldgas_alt}, but at the expense of a slightly
larger cooled gas mass at higher redshifts. Adopting a smaller core
radius (e.g. $r_{\rm c}=0.15r_{\rm s}$) makes very little difference.
Finally, we note that assuming a gas profile which traces that of the
dark matter leads to $\sim 50\%$ more cooling in the highest mass
halos (note this model is not shown in Fig.~\ref{fig:coldgas_alt}).

\begin{figure*}
\hspace{5mm}\psfig{file=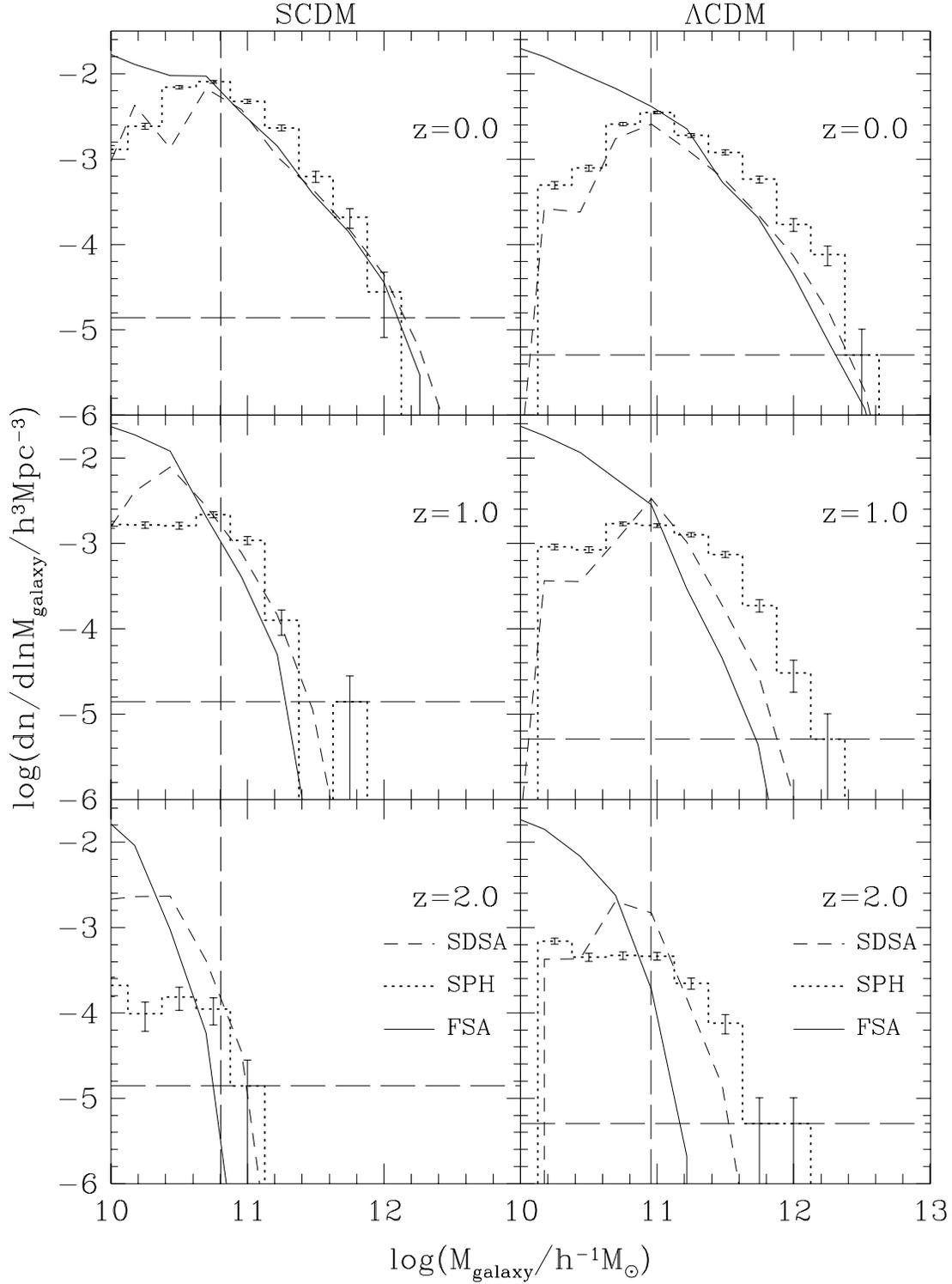,width=160mm} \caption{The
differential galaxy mass function in the SCDM (left-hand panels)
and $\Lambda$CDM (right-hand panels) cosmologies at various
epochs. The solid line corresponds to the \FSA\ model, the dotted
histogram to the SPH simulation and the dashed line to the \SDSA\
model. The vertical dashed line indicates the critical mass for
cooling in the SPH simulations (which is 64 times the gas particle
mass), whilst the horizontal dashed line indicates the abundance
corresponding to one object in the entire SPH simulation volume.
Results are shown at $z=0$, 1 and 2 as indicated in each panel.}
\label{fig:cgdiff}
\end{figure*}

\begin{figure}
\psfig{file=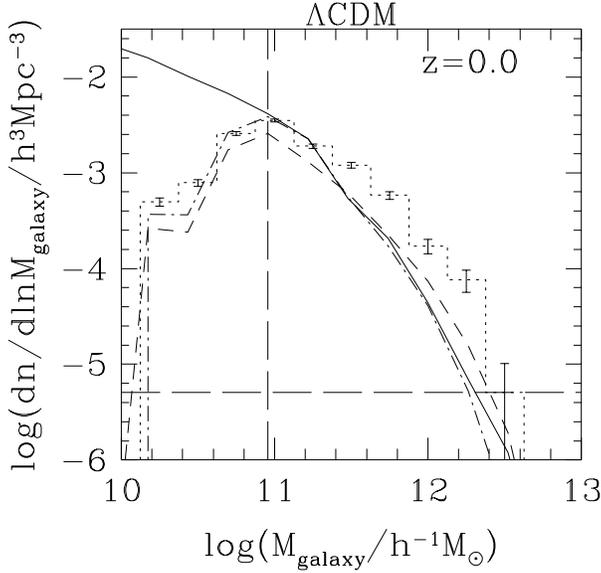,width=80mm,bbllx=5mm,bblly=160mm,bburx=110mm,bbury=270mm,clip=}
\caption{The differential galaxy mass function at $z=0$ in the
$\Lambda$CDM cosmology. The solid line corresponds to the \FSA\
model, the dotted histogram to the SPH simulation and the dashed
line to the \SDSA\ model. Dot-dashed lines show the results for
the \SDSA\ model using the standard Press-Schechter dark matter
halo mass function and the standard extended Press-Schechter
formalism (i.e. with $f_{\delta_{\rm c}}=1$) for generating merger
trees. The vertical dashed line indicates the critical mass for
cooling in the SPH simulations (which is 64 times the gas particle
mass), whilst the horizontal dashed line indicates the abundance
corresponding to one object in the entire SPH simulation volume.}
\label{fig:mfuncorr}
\end{figure}

\begin{figure*}
\psfig{file=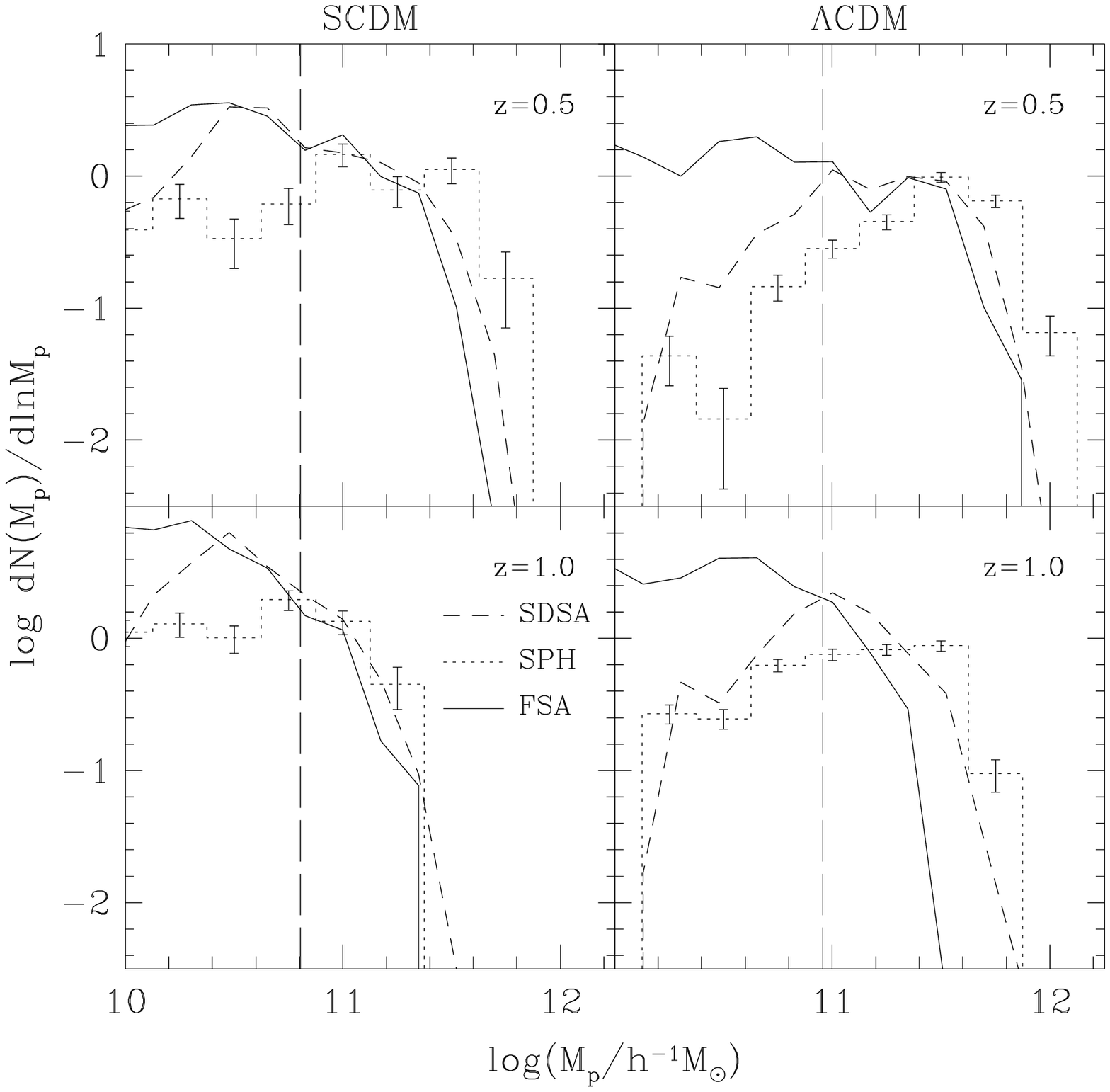,width=160mm,bbllx=0mm,bblly=85mm,bburx=190mm,bbury=270mm,clip=}
\caption{The mass function of galaxies which are progenitors of
present-day galaxies with (cold gas plus stellar) masses
$10^{11.5}$--$10^{12}h^{-1}{\rm M}_\odot$. Results are shown at
$z=0.5$ and 1 for SCDM (left-hand panels) and $\Lambda$CDM (right-hand
panels). The solid line corresponds to the \FSA\ model, the dotted
histogram to the SPH simulation and the dashed line to the \SDSA\
model. The vertical dashed line indicates the mass corresponding to 64
SPH gas particles.} \label{fig:galprogen}
\end{figure*}

\subsection{Properties of galaxies}
\label{sec:galgas}

We now consider properties of the models that are sensitive to the way
in which gas is apportioned amongst individual galaxies, rather than
just the total amount of gas in a halo. In the SPH simulation, the
number of galaxies that form in a given halo and their masses are
determined by resolution effects, the gas cooling rate and the galaxy
merger rate. In the semi-analytic models, the choice of density
profile for the hot corona determines the cooling rate of the gas. We
also expect the masses and numbers of galaxies to be affected by the
choice of $f_{\rm df}$, which controls the rate at which galaxies
merge within a halo and, in the case of SDSA model, the choice of
$N'_{\rm SPH}$, which determines the minimum halo mass in which
galaxies can form.
 
\subsubsection{Galaxy mass function}
\label{sec:massfunc}
 
Figure \ref{fig:cgdiff} shows the differential galaxy mass functions
in the models at $z=0$, 1 and 2. (Recall that, by definition, $M_{\rm
galaxy}$ consists only of cold gas in the SPH and \SDSA\ cases, but
can also include stars in a disk and spheroid in the \FSA\ case.)  The
agreement between the SPH and semi-analytic models is somewhat
different in the two cosmologies. In the SCDM case, there is quite
good agreement for large galaxy masses, but the SPH simulation
produced about twice as many galaxies with mass of a few times
$10^{11}M_{\odot}$ than the \SDSA\ model. In the $\Lambda$CDM case,
the SPH simulation produced about 3--4 times as many galaxies as the
\SDSA\ model over most of the mass range above the critical cooling
mass. If viewed as a difference in mass at fixed abundance, the
discrepancy can be as large as a factor of almost 2. For a fixed
mass-to-light ratio, this would lead to the bright end of the
luminosity function being approximately 0.75 magnitudes brighter in
the SPH simulation than in the \SDSA\ model. The differences between
\SDSA\ and SPH models are similar at higher redshifts, whilst the
\FSA\ mass function evolves more rapidly.  We discuss the possible
reasons for these discrepancies shortly.
 
First, we note the interesting behaviour at the low mass end of the
mass function. Below the critical cooling mass, $9.0 \times 10^{10}$
and $6.4 \times 10^{10}h^{-1} M_{\odot}$ in $\Lambda$CDM and SCDM
respectively, the abundance of galaxies in the \SDSA\ and SPH models
drops very sharply, due to the effects of resolution. These prevent
the formation of small halos and also inhibit the cooling of gas in
marginally resolved halos. In the \FSA\ model, on the other hand, the
number of galaxies continues to rise towards small masses. Although
feedback becomes gradually more efficient towards lower masses, 
this still allows many low mass galaxies to form.  In
both the \SDSA\ and SPH models some galaxies do form with mass
below the cooling limit. In the former case this can occur because
cooling is prevented only in halos with a total hot gas mass lower
than the limit.  It is still possible for cooling in a more massive
halo to create a galaxy below the mass threshold simply because there
has not been enough time for the total hot gas mass of the halo to
cool. The SPH simulations also contain galaxies below the mass limit
because the cutoff is not sharp, but marks the point at which cooling
becomes efficient.  These objects represent young galaxies in the
process of formation and, except at early times, they make up a small
fraction of the total number of galaxies. In addition, in the SPH
model there is a small contribution at low masses from galaxies that
are being tidally disrupted in clusters.

In Fig. \ref{fig:mfuncorr} we show the effect on the \SDSA\ galaxy mass
function (in the $\Lambda$CDM cosmology) of using the standard
Press-Schechter halo mass function and $f_{\delta_{\rm c}}=1$, rather than
the Sheth-Mo-Tormen halo mass function and $f_{\delta_{\rm c}}$ as given by
eqn.~(\ref{eq:fdc}). Using the latter, more accurate, mass function
produces somewhat more massive galaxies.  The choice of $f_{\delta_{\rm
c}}$ alters the mass function of progenitors of the more massive galaxies,
but has little effect elsewhere.

Some insight into the origin of the differences between the galaxy mass
functions in the SPH and \SDSA\ models may be obtained by considering the
mass function of galaxy progenitors.  For present-day galaxies with masses
in the range $10^{11.5}$--$10^{12}h^{-1}{\rm M}_\odot$, Fig.~\ref{fig:galprogen}
shows progenitor mass functions at $z=0.5$ and $z=1$. These are normalized
to the present day abundance of galaxies (i.e. $({\rm d}N(M_{\rm P})/{\rm
d}\ln M_{\rm P}){\rm d}\ln M_{\rm P}$ is the number of progenitors in the
mass range $\ln M_{\rm P} $ to $\ln M_{\rm P} +{\rm d}\ln M_{\rm P}$ per
parent galaxy) so that the differences in the $z=0$ mass functions seen in
Fig. \ref{fig:cgdiff} do not propagate through to this comparison. It is
immediately obvious from this figure that the \SDSA\ galaxies have fewer
high-mass progenitors than the galaxies in the SPH simulation, particularly
in the $\Lambda$CDM cosmology.

The discrepancy between the SPH and semi-analytic galaxy mass
functions seen in Figs.~\ref{fig:cgdiff} and~~\ref{fig:galprogen}
could be due either to differences in the galaxy merger rates or in
the gas cooling rates in the two models.  We first consider the
possibility that the SPH and \SDSA\ galaxy merger rates may be
different, leading to different numbers of galaxies forming in each
halo.  As we discussed in \S\ref{sec:SPHass}, merging in the SPH
simulations may be artificially enhanced by non-physical effects
introduced, for example, by the use of softened forces. The strength
of these effects will depend in a non-trivial way on the formation
epoch of each galaxy. Whilst we cannot attempt to mimic the details of
such effects in the semi-analytic model, we can explore the
consequences of a global change in the merger timescale. In the
semi-analytic model, when a galaxy falls into a larger halo, it is
assumed to sink to the centre in a time proportional to the dynamical
friction timescale for an object orbiting in an isothermal halo. This
may be written as \cite{lc93}:
\begin{equation} \tau _{\mathrm mrg} = f_{\mathrm df} \theta _{\rm orb} \tau _{\mathrm dyn} {0.3722 \over \ln (\Lambda _{\mathrm Coulomb})} {M_{\mathrm H} \over M_{\mathrm sat}},
\label{eq:merge}
\end{equation}
where $f_{\mathrm df}$ is a dimensionless parameter; $M_{\mathrm H}$
is the mass of the halo in which the satellite orbits; $M_{\mathrm
sat}$ is the mass of the satellite galaxy including the mass of the
dark matter halo in which it formed (not including the mass of the
satellite's dark matter halo leads to an overestimate of the dynamical
friction timescale as shown by \pcite{jfn95}); $\tau _{\mathrm dyn}$
is the dynamical time of the large halo and the Coulomb logarithm is
$\ln (\Lambda _{\mathrm Coulomb}) = \ln (M_{\mathrm H}/M_{\mathrm
sat})$. The variable $\theta _{\rm orb}$ contains the dependence of
the merger timescale on the orbital parameters of the galaxy and is
chosen from the distribution found in N-body simulations by
\scite{tormen97}, as described by \scite{coleetal99}.

\scite{coleetal99} point out that the formula in
eqn.~(\ref{eq:merge}) has been derived on the basis of a number of
assumptions, for example that the galaxy may be treated as a point
mass. Thus, whilst they recommend a default value of $f_{\rm df}=1$,
they also allow themselves the freedom to choose a different value if
required to produce a realistic model. Recently, \scite{colpi99}
have examined the validity of eqn.~\ref{eq:merge} in detail using both
analytic and numerical techniques. On the basis of their
investigations, they suggest small modifications to the formula, in
particular a slightly different dependence of $\theta_{\rm orb}$ on
the orbital eccentricity. They also find that the effects of tidal
stripping produce $f_{\rm df}\approx 2.7$ (for a specific model of the
satellite's dark matter halo). Both of these changes act to increase
$\tau_{\rm mrg}$, and so reduce the merger rate. Since the \SDSA\
model already contains too few high mass galaxies compared to the SPH
simulations, a slower merger rate would merely increase the
discrepancy.

At the resolution of our SPH simulations, an infalling satellite
galaxy will lose nearly all of its original dark matter halo shortly
after entering the larger halo. On the other hand, in the SPH
simulations, merging with the central object is driven not only by
dynamical friction, but also by drag due to viscous effects as the
satellite moves through the hot halo of the cluster
\cite{fews,tittley99}. These processes may drive the effective $f_{\rm
df}$ to a value less than unity. 

\begin{figure}
\psfig{file=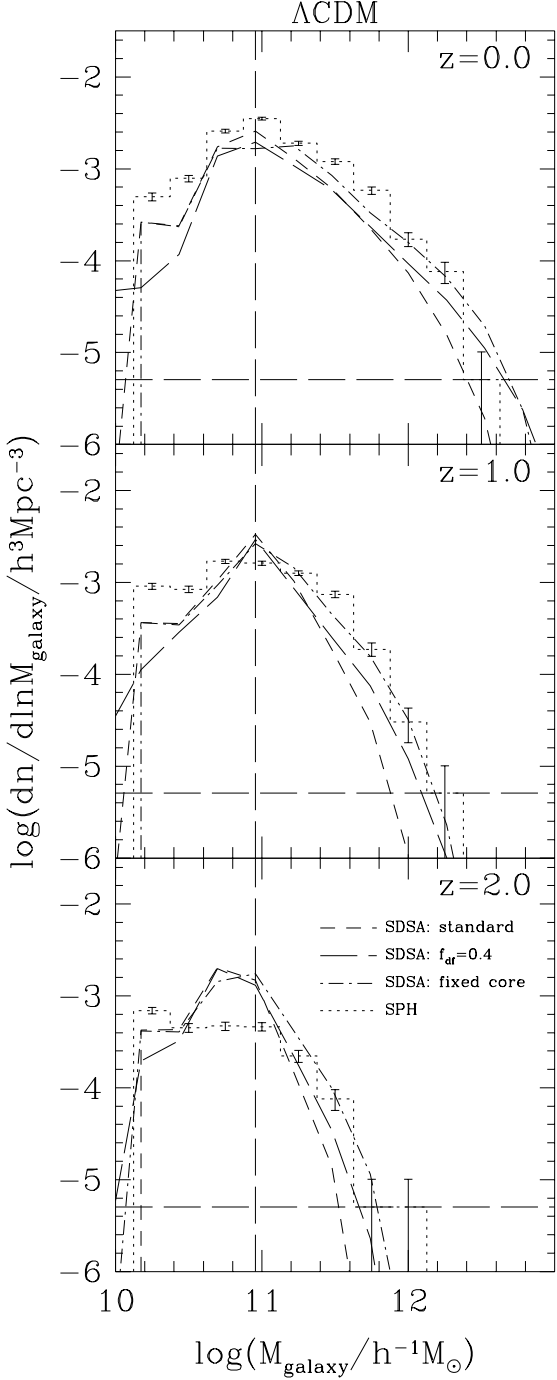,width=80mm,bbllx=0mm,bblly=0mm,bburx=110mm,bbury=270mm,clip=}
\caption{The differential galaxy mass function at selected redshifts
in the $\Lambda$CDM cosmology predicted by the \SDSA\ model when (i)
using a faster merging rate (long-dashed line) and (ii) keeping the
core radius of the hot gas density profile fixed (dot-dashed
line). For reference, the short-dashed line shows the standard \SDSA\
model. Dotted histograms show the SPH simulation results.  The
vertical dashed lines indicate the critical mass for cooling in the
SPH simulations (which is 64 times the gas particle mass), whilst the
horizontal dashed line indicates the abundance corresponding to one
object in the entire SPH simulation volume.  Results are shown at
$z=0$, 1 and 2, as indicated in each panel.}
\label{fig:cg_altmrg}
\end{figure}
 
The smaller the value of $f_{\mathrm df}$, the faster a galaxy will
sink to the centre of its host halo. In Fig. \ref{fig:cg_altmrg} we
show the effect of reducing the merging rate from the default value,
$f_{\mathrm df}=1$, to $f_{\mathrm df} = 0.4$ on the \SDSA\ galaxy
mass function in the $\Lambda$CDM model. This change improves the
match between the \SDSA\ and SPH models at $z=0$, but only slightly,
and mostly at high masses. Below $10^{12}h^{-1}{\rm M}_\odot$, the two
models still differ significantly. At higher redshift the improvement
is slightly better, but the \SDSA\ model still contains fewer
high-mass objects than the SPH simulation.  Thus, a change in the
global merger timescale does not help to reconcile the two
models. This is perhaps not too surprising since this simple
alteration cannot mimic any redshift dependence of the merger
timescales in the SPH simulation. Decreasing $f_{\rm df}$ further
produces more high mass galaxies, at the expense of depleting the
number of lower mass galaxies, thereby worsening the agreement between
the \SDSA\ and SPH galaxy mass functions at the low mass end.

A second possible explanation for the discrepancy between the galaxy
mass functions is that the relative gas cooling rates in the SPH and
\SDSA\ models are different.  We investigate the effects of changing
the rate at which gas cools in the \SDSA\ model, by keeping the core
radius of the hot gas density profile equal to a fixed fraction of the
NFW scale radius in the halo.  This has the effect of allowing gas to
cool more rapidly in any given halo.  The galaxy mass function of the
\SDSA\ model for this case is shown by the dot-dashed line in
Fig. \ref{fig:cg_altmrg}. The agreement with the SPH simulation is now
much improved.

In the next section we will explore how changing merger and cooling
rates affects the number of galaxies found in halos of a given mass
and thus identify the underlying cause of the differences between the
mass functions in the SDSA and SPH models.
 
\subsubsection{Number of galaxies per halo}

\begin{figure}
\psfig{file=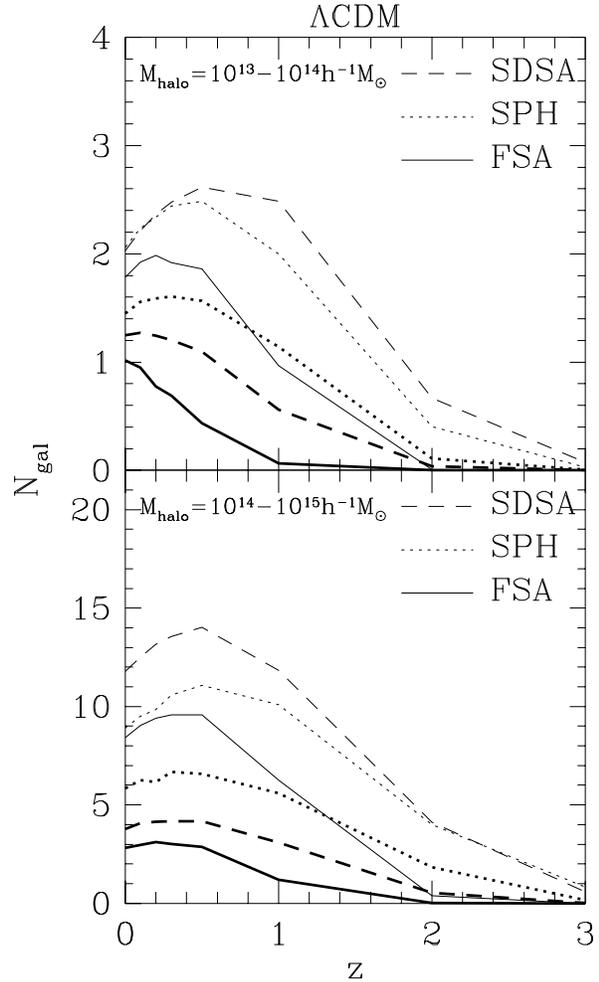,width=80mm,bbllx=5mm,bblly=90mm,bburx=110mm,bbury=270mm,clip=}
\caption{The mean number of progenitor galaxies with mass greater than
64 SPH particles (thin lines) and 128 particles (heavy lines) in the
$\Lambda$CDM cosmology, as a function of redshift, for halos of
present-day mass $10^{13}$--$10^{14}h^{-1}{\rm M}_\odot$ (upper panel)
and $10^{14}$--$10^{15}h^{-1}{\rm M}_\odot$ (lower panel). Dashed,
dotted and solid lines show results from the \SDSA, SPH and \FSA\
models respectively.}
\label{fig:haloNgal}
\end{figure}

The redshift evolution of the mean number of progenitor galaxies per
halo in the $\Lambda$CDM cosmology is shown in
Fig.~\ref{fig:haloNgal}, for two ranges of present-day halo masses:
$10^{13}$--$10^{14}h^{-1}{\rm M}_\odot$ (upper panel) and
$10^{14}$--$10^{15}h^{-1}{\rm M}_\odot$ (lower panel). (Qualitatively
similar results are obtained for the SCDM cosmology.) In the SPH
simulation there are 165 halos in the lower mass range and 12 in the
higher mass range. Since the galaxy abundance is a steep function of
galaxy mass (c.f. Fig.~\ref{fig:cgdiff}) the number of galaxies per
halo above some particular mass cut is always dominated by galaxies
close to the cutoff. As a result, the number of galaxies per halo
more massive than 64 SPH particles may be affected by differences in
the way in which the resolution limit works in SPH and \SDSA\
models. Therefore, we plot results for galaxies more massive than 64
SPH gas particles (thin lines), but also for those more massive than
128 SPH gas particles (heavy lines) which are much less affected by
resolution effects.

The shape of the abundance curves in Fig.~\ref{fig:haloNgal} is
determined by the relative rates of galaxy formation and merging and
has the same basic form in the \SDSA\ and SPH models. The number of
galaxies more massive than 64 SPH gas particles predicted by the
\SDSA\ model in halos of mass $10^{13}$--$10^{14}h^{-1}{\rm M}_\odot$
at the present day agrees well with that found in the SPH simulation.
The level of agreement, however, is less impressive at higher redshift
and for higher mass halos. For the 128 particle selection the \SDSA\
model contains
\emph{fewer} galaxies per halo than the SPH model in both ranges of
halo mass and at all redshifts.

\begin{figure}
\psfig{file=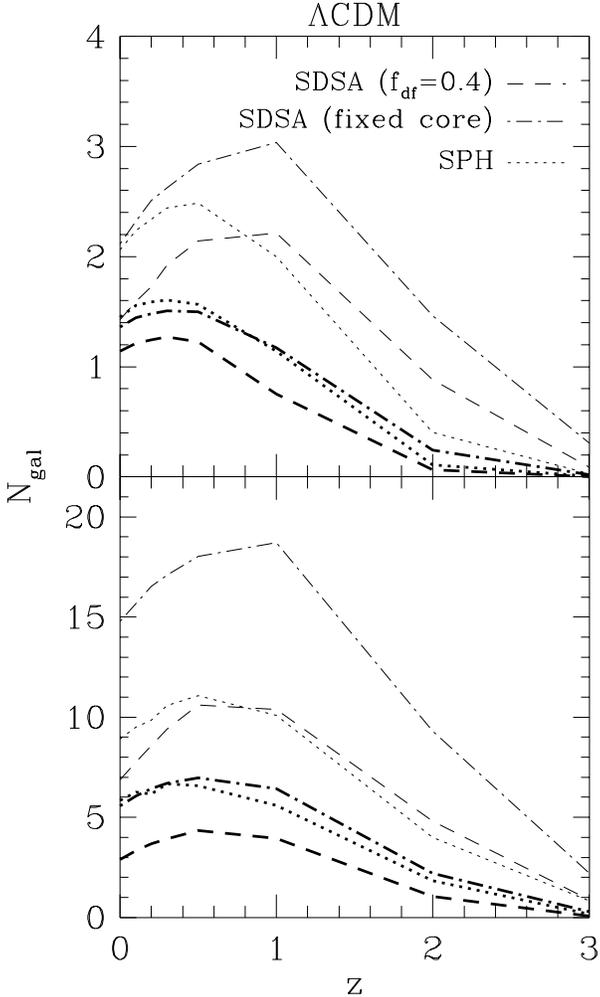,width=80mm,bbllx=85mm,bblly=90mm,bburx=188mm,bbury=270mm,clip=}
\caption{ The mean number of progenitor galaxies with mass greater
than 64 SPH particles (thin lines) and 128 particles (heavy lines), as
a function of redshift in halos of present-day mass
$10^{13}$--$10^{14}h^{-1}{\rm M}_\odot$ (upper panel) and
$10^{14}$--$10^{15}h^{-1}{\rm M}_\odot$ (lower panel) in the
$\Lambda$CDM cosmology.  The predictions of the \SDSA\ model are shown
in the cases where (i) a faster merger rate is used (dashed line) and
(ii) the cooling rate is enhanced by keeping the core radius fixed in
the hot gas density profile (dot-dashed line). The dotted line shows
the results from the SPH simulation.}
\label{fig:Ngal_altmrgbig}
\end{figure}

In the preceeding subsection, we saw that increasing the merger rate
in the $\Lambda$CDM \SDSA\ model leads to slightly better agreement
between its galaxy mass function and that of the SPH simulation. A
faster merger rate, however, depletes the number of galaxies by
combining them into larger galaxies. In Fig.~\ref{fig:Ngal_altmrgbig},
we show the effect of reducing $f_{\mathrm df}$ from 1 to 0.4 on the
evolution of the progenitor population in the \SDSA\ model. For
galaxies more massive than 64 SPH gas particles, the rate at which
$N_{\rm gal}$ increases at high redshift is now slower than before and
results in a lower value of $N_{\rm gal}$ at the present day. The net
effect is to bring the \SDSA\ curve closer to the SPH results for the
$10^{14}$--$10^{15}h^{-1}{\rm M}_\odot$ halo sample, but to increase
the discrepancy for the $10^{13}$--$10^{14}h^{-1}{\rm M}_\odot$
halos. For galaxies more massive than 128 SPH gas particles, the
already low galaxy numbers are depleted even further, exacerbating the
discrepancy with the SPH results. We conclude that the discrepancies
in the mass functions in the $\Lambda$CDM \SDSA\ and SPH models are
not due to differences in the galaxy merger rates in the two models.

The effects of altering the cooling rate in the \SDSA\ model are also
illustrated in Fig.~\ref{fig:Ngal_altmrgbig} (dot-dashed lines). As
before, we have varied the cooling rate simply by keeping the core
radius of the hot gas density profile fixed. For galaxies more massive
than 64 SPH gas particles, the \SDSA\ model now overpredicts the mean
number of galaxies per halo at all redshifts, particularly in the most
massive halos. The increase is driven by galaxies near the 64 particle
cutoff whose mass increases due to the additional cooling of gas.  (A
similar behaviour occurs if the cooling rate is enhanced by reducing
the threshold for cooling from $N'_{\rm SPH}=64$ to 32, rather than by
changing the density profile of the gas.)

A cleaner test can be made by looking at galaxies more massive than
128 SPH gas particles. In this regime, the \SDSA\ model with enhanced
cooling is in excellent agreement with the SPH simulation. As we saw
earlier, such a modification of the \SDSA\ model also produces a
galaxy mass function (Fig.~\ref{fig:cg_altmrg}) and a distribution of
cold gas in halos of different mass (Fig.~\ref{fig:coldgas_alt}) that
are very similar to those in the SPH simulation. We conclude therefore
that the main reason for the differences we have found between the
\SDSA\ model and the SPH simulation is that gas cools more efficiently
in massive halos in the SPH simulation. This difference, and the
corresponding differences in the mass function, are larger in
$\Lambda$CDM than in SCDM, pressumably because of the larger time
interval during which gas can cool in $\Lambda$CDM. Increasing the
cooling rate in the \SDSA\ model slightly spoils the excellent
agreement with the SPH simulation on the global phase fractions
(Fig.~\ref{fig:gastemp}). The effect, however, is small (a maximum
difference of 50\% as opposed to the original 25\%) and of the same
order as the effect of assuming a different resolution limit for the
\SDSA\ model (Fig.~\ref{fig:gastemp_NSPH32}).

\subsection{Spatial distribution of galaxies}
\label{sec:xi}
 
\begin{figure*}
\psfig{file=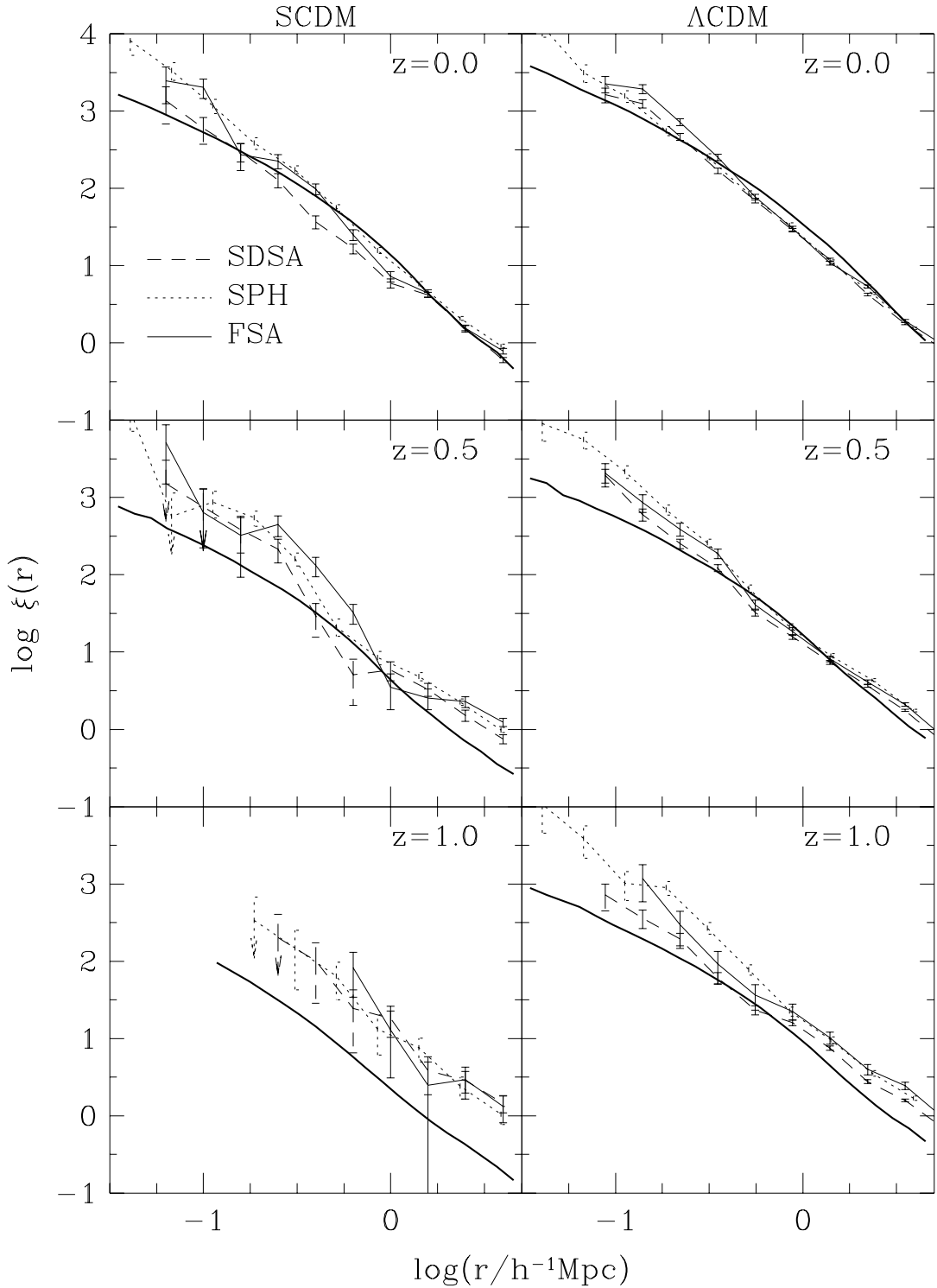,width=160mm}
\caption{Two-point galaxy correlation functions from the SPH, \FSA\
and \SDSA\ models are shown as the dotted, solid and dashed lines
respectively. The thick solid line shows the correlation function of dark
matter in the simulations. Galaxies from all the models are chosen to
have a mass in cold gas plus stars greater than 64 times the SPH gas
particle mass. SCDM and $\Lambda$CDM cosmologies are shown in left and
right-hand panels respectively. Results are shown at $z=0$, 0.5 and
1 as indicated in each panel. Downwards-pointing arrows are used
where the lower errorbar drops below zero.}
\label{fig:xi}
\end{figure*}

As a final comparison, we consider the clustering of galaxies, as
measured by the two-point correlation function. This is plotted in
Fig.~\ref{fig:xi} for galaxies more massive than $N'_{\rm SPH}$ gas
particles. Note that this selection criterion (which picks out only
rather massive galaxies) is very different from that considered by
\scite{ajbclust} and, as a result, the correlation functions plotted
in Fig.~\ref{fig:xi} are quite different from those in
\scite{ajbclust}.  Note also that the relatively small volume of the
simulations affects the determination of the correlation function for
pair separations greater than a few Mpc.  To compute the correlation
function in the \SDSA\ and \FSA\ models, we make the further
assumption that the galaxies trace the mass within each dark matter
halo.

The agreement of the correlation functions of all three models in both
cosmologies is very good, except on scales below $1h^{-1}$Mpc in the
SCDM cosmology, where the semi-analytic models have a lower amplitude
than the SPH simulation. The good agreement is not completely
unexpected as the mean number of galaxies per halo is similar in all
models, but it does demonstrate that the assumption that galaxies
trace the mass within individual halos is a reasonable approximation,
at least for studies of the galaxy correlation function. In
particular, all the models predict very similar evolution in the
correlation function, with galaxies becoming strongly biased at high
redshifts.

\begin{figure}
\psfig{file=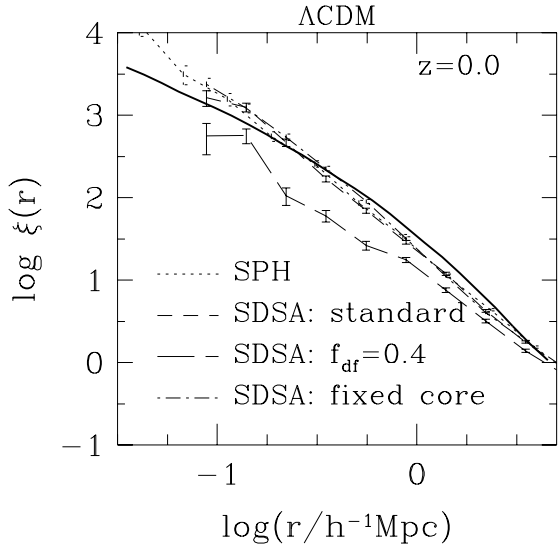,width=80mm,bbllx=0mm,bblly=165mm,bburx=110mm,bbury=270mm,clip=}
\caption{The $\Lambda$CDM galaxy two-point correlation function in the
SPH and \SDSA\ models at $z=0$. The SPH model is shown by the dotted
line, whilst the standard \SDSA\ model is indicated by the
short-dashed line. The long-dashed line shows the \SDSA\ model with an
enhanced merger rate ($f_{\rm df}=0.4$), whilst the dot-dashed line
shows the \SDSA\ model with a fixed gas core radius. The heavy solid
line shows the correlation function of dark matter.}
\label{fig:xi_altmrgbig}
\end{figure}

In Fig.~\ref{fig:xi_altmrgbig} we show the effects of reducing the
merger timescale in the $\Lambda$CDM cosmology on the galaxy two-point
correlation function. The enhanced merger rate significantly lowers
the \SDSA\ correlation function below that of the SPH model. Enhancing
the cooling rate of gas (by using a fixed gas core radius) in the
\SDSA\ model has a negligible effect on the correlation function. This 
is further evidence that the main difference between the \SDSA\ and
SPH models is not a gross difference in galaxy merger rates, but
rather a difference in the efficiency with which gas cools. 

\section{Discussion and conclusions}
\label{sec:discuss}
 
In this paper we have explored the consistency of two very different
techniques commonly used to study galaxy formation, semi-analytic
modelling and SPH simulations. For this purpose, we constructed a
simplified or `stripped-down' semi-analytic model (SDSA), designed to
mimic the conditions of the SPH simulations as closely as possible. An
artificial resolution limit, similar to that in our SPH simulations,
was imposed on the \SDSA\ model and the standard star formation and
feedback prescriptions normally used in semi-analytic modelling were
stripped out. Furthermore, the dark matter halo mass function proposed
by \scite{smt} was used in the \SDSA\ model instead of the
standard Press-Schechter formula since the former provides a better
match to the mass function in N-body simulations. Similarly, we also
modified the extended Press-Schechter theory in order to better match
the distribution of progenitor halo masses obtained in the SPH
simulations. In this way, we minimize differences in the evolution of
dark matter halos and concentrate on the aspect of interest in this
work: the evolution of the gas. In this paper, we have focussed on a
{\it statistical} analysis of the differences between the SDSA and SPH
models. As a secondary aim, we have also compared the outcome of the
SPH simulations with a full semi-analytic model, in which the star
formation and feedback prescriptions are restored and no artificial
resolution limit is imposed. The motivation for this latter comparison
is a desire to assess how the realism of a typical SPH simulation is
likely to be compromised by the neglect, or very rough treatment, of
star formation and feedback. 
 
For our main comparisons, we considered five properties of the gas
distribution in two versions of the CDM cosmology (SCDM and
$\Lambda$CDM), over the redshift range 0 to 2. In order of decreasing
generality, these are: (i) the global fractions of gas in three
distinct phases --- hot halo, galactic (i.e. cold and dense gas) and
uncollapsed; (ii) the amount of cold galactic gas in halos of
different mass; (iii) the abundance of objects as a function of their
cold gas mass; (iv) the number of ``galaxies'' as a function of halo
mass; and (v) the correlation function of galaxies.  The main
conclusion of this paper is that the agreement between the SPH
simulation and the stripped-down version of the semi-analytic model is
better than a pessimist might have expected. The tests we have carried
out generally show reasonable agreement between the two techniques,
although we have found significant differences in the cold gas mass
functions in the two models.  Since in both approaches the gas physics
are necessarily treated in an approximate fashion, it seems
inappropriate to regard one as `correct' and the other as `wrong,' in
the instances where discrepancies arise.

Specifically, our main results may be summarized as follows:
 
(i) Over the simulation volume as a whole, the global amounts of gas
in the three main phases that develop (hot, cold galactic and
uncollapsed) are very similar in the \SDSA\ model and the SPH
simulation. From $z=0.5$ (the redshift by which about half the final
amount of galaxy gas has cooled) to the present, the maximum
difference is never larger than 25\%. Perhaps surprisingly, the \SDSA\
model produces slightly {\it less} hot gas than the SPH simulation at
all epochs (although this difference may not be significant given the
uncertainties inherent in this comparison). The distribution of cold
gas in halos of different mass also agrees quite well (to within $\sim
50\%$) with the largest differences occuring in the most massive
halos. These results apply to the SPH and \SDSA\ models in their
standard forms, i.e. with parameters chosen independently of this
comparison in order to match certain observational constraints.
Thus, we conclude that in spite of its various approximations
(e.g. spherical symmetry), the simple model of gas cooling normally
employed in semi-analytic models gives similar results, on average, 
to the SPH simulations. 

(ii) Even though the total amount of cold gas is similar in the
\SDSA\ model and the SPH simulations, the actual cold gas mass
functions (i.e. the abundance of galaxies as a function of cold gas mass)
are different, particularly in the $\Lambda$CDM cosmology. In this case,
the SPH simulation produced three to four times more high-mass galaxies
than the \SDSA\ model, or equivalantly, at a fixed abundance, galaxies in
the SPH simulation are, on average, twice as massive as their \SDSA\
counterparts. We were able to discount a difference in galaxy merging rates
as the dominant source of this discrepancy. Instead, we identified as the
culprit a difference in the efficiency of cooling in massive galaxies: more
gas cools into these galaxies in the SPH simulation than in the standard
\SDSA\ model. Thus, artificially increasing the cooling rate of gas in the
\SDSA\ model (by 
assuming that the core radius of the gas density profile remains fixed
rather than growing with time as in our standard model) leads to excellent
agreement with the SPH galaxy mass functions and also with the mass
functions of galaxy progenitors.

(iii) The spatial distribution of galaxies, as characterised by the
two-point correlation function, is remarkably similar in the semi-analytic
and SPH models. This is true not only of the present-day distribution, but
also of the clustering at high redshift. This conclusion also holds
even when the cooling rate is enhanced in the \SDSA\ model as
discussed in (ii). 

The primary limitation of current cosmological SPH simulations is the
relatively poor resolution attainable even with the largest computers.
By comparing our SPH or \SDSA\ models to the full semi-analytic model
(FSA), we gain some idea of how important these resolution effects are
in practice. Furthermore, since the \FSA\ model includes prescriptions
for star formation and feedback that are not modelled at all in the
SPH simulations we have considered, we can also assess how important
these processes are in determining the properties of hot and cold gas
in the moderately large galaxies that form in our SPH simulations.
The total amount of gas that can cool in an SPH simulation is
determined by the resolution limit.  On the other hand, because of its
intrinsically high resolution, the semi-analytic model traces the
evolution of gas even in small halos which the SPH simulation assigns
to the uncollapsed phase. As a result, not only is the fraction of
halo gas (hot and cold) larger in the \FSA\ model than in the SPH
simulation, but its cold gas mass function extends to smaller masses
that can be resolved in the SPH simulation. The FSA cold gas mass
function has a much sharper cut-off at the massive end than either the
\SDSA\ or SPH models.

In summary, our comparisons demonstrate a higher level of consistency
than was perhaps expected between the results of SPH simulations and
the more idealized semi-analytic models. A particularly uncertain
component of the semi-analytic treatment is the assumption that gas
cools from a quasi-equilibrium state established when the gas is
shock-heated to the virial temperature of a halo during collapse. Our
comparisons do not test this assumption directly, only its net effect
on the amount of gas that cools. Globally, this turns out to be very
similar to the SPH result. However, the semi-analytic model tends to
produce somewhat less cool gas in massive halos than the SPH
simulation, particularly in the $\Lambda$CDM cosmology.  We stress
that due to the limited resolution of our SPH simulations, our
conclusions are restricted to massive galaxies, with baryonic mass of
$\gsim 10^{11}M_{\odot}$. It will be important to check whether our
results still hold in higher resolution simulations.  In this paper we
have focussed on statistical properties of the galaxy
population. Agreement at this level does not necessarily imply
agreement on the properties of galaxies on a case-by-case basis. We
intend to examine this question in future work. Our present results,
however, provide useful support for the reliability of current
techniques for modelling galaxy formation in a cosmological context.

\section*{Acknowledgements}

AJB and CSF acknowledge receipt of a PPARC Studentship and Senior
Fellowship respectively. CSF also acknowledges receipt of a Leverhulme
Research Fellowship. This work was supported in part by a PPARC
rolling grant and by the European Community's TMR Network for Galaxy
Formation and Evolution.  We acknowledge the Virgo Consortium for
making available the simulations used in this study and our
colleagues, Shaun Cole and Cedric Lacey, for allowing us to use
results from the Durham semi-analytic galaxy formation model. We are
grateful to Shaun Cole and David Weinberg for many stimulating
discussions, and to James Binney whose incisive questions encouraged
us to pursue this project. These simulations were carried out on the
T3D at the Edinburgh Parallel Computing Centre (EPCC).

\end{document}